\documentclass[12pt]{article}
\usepackage{epsfig,amsfonts,amssymb}
\usepackage{hyperref}
\usepackage{subcaption}

\pdfoutput=1

\usepackage{cite}
\usepackage{cite}

\usepackage{comment}

\def\ZZZ{{\hbox{ Z\kern-1.6mm Z}}}
\def\RRR{{\hbox{ R\kern-2.4mm R}}}
\def\CCC{{\hbox{ C\kern-2.0mm C}}}
\def\zzz{{\hbox{z\kern-1mm z}}}

\newcommand{\vt}{\vartheta}

\newcommand{\qeq}{{\hbox{=\kern-2.3mm ? \kern.5mm }}}
\renewcommand{\qeq}{=}

\newcommand{\eps}{\epsilon}

\newcommand{\BB}{{\cal B}}

\newcommand{\FF}{{\cal F}}

\newcommand{\MM}{{\cal M}}

\newcommand{\OO}{{\cal O}}

\newcommand{\wt}{\widetilde}

\newcommand{\be}{\begin{equation}}
\newcommand{\ee}{\end{equation}}
\newcommand{\ben}{\begin{eqnarray}\displaystyle}
\newcommand{\een}{\end{eqnarray}}

\newcommand{\refb}[1]{(\ref{#1})}
\newcommand{\p}{\partial}
\newcommand{\sectiono}[1]{\section{#1}\setcounter{equation}{0}}

\def\one{{\hbox{ 1\kern-.8mm l}}}
\def\zero{{\hbox{ 0\kern-1.5mm 0}}}

\newcommand{\bea}[1]{\begin{eqnarray}\label{#1} }
\newcommand{\eea}{\end{eqnarray}}

\newcommand{\eqref}{\refb}

\usepackage{bm}
\usepackage[table]{xcolor}

\def\erfi{{\rm erfi}}

\def\figone{

\def\JPicScale{0.8}
\ifx\JPicScale\undefined\def\JPicScale{1}\fi
\unitlength \JPicScale mm
\begin{picture}(110,70)(0,0)
\linethickness{0.5mm}
\multiput(10,70)(0.12,-0.36){83}{\line(0,-1){0.36}}
\linethickness{0.5mm}
\multiput(10,10)(0.12,0.36){83}{\line(0,1){0.36}}
\linethickness{0.5mm}
\put(50,10){\line(0,1){60}}
\linethickness{0.5mm}
\multiput(80,10)(0.12,0.36){83}{\line(0,1){0.36}}
\linethickness{0.5mm}
\multiput(80,70)(0.12,-0.36){83}{\line(0,-1){0.36}}
\linethickness{0.2mm}
\qbezier(50,60)(60.5,49.62)(60.5,40)
\qbezier(60.5,40)(60.5,30.38)(50,20)
\linethickness{0.2mm}
\qbezier(90,40)(100.44,50.5)(105.25,50.5)
\qbezier(105.25,50.5)(110.06,50.5)(110,40)
\qbezier(110,40)(110.06,29.5)(105.25,29.5)
\qbezier(105.25,29.5)(100.44,29.5)(90,40)
\put(10,0){\makebox(0,0)[cc]{(a)}}

\put(50,0){\makebox(0,0)[cc]{(b)}}

\put(95,0){\makebox(0,0)[cc]{(c)}}

\end{picture}

}

\def\figsix{

\ifx\JPicScale\undefined\def\JPicScale{1}\fi
\unitlength \JPicScale mm
\begin{picture}(140,55.59)(0,0)
\linethickness{0.2mm}
\put(45.6,50){\circle{11.18}}

\linethickness{1mm}
\put(20,50){\line(1,0){10}}
\linethickness{0.2mm}
\put(30,50){\line(1,0){10}}

\linethickness{1mm}
\put(60,50){\line(1,0){10}}
\linethickness{0.2mm}
\put(70,50){\line(1,0){10}}

\linethickness{1mm}
\put(90,50){\line(1,0){10}}
\linethickness{0.2mm}
\put(105.5,50){\circle{11.18}}

\linethickness{1mm}
\put(125,50){\line(1,0){15}}

\put(30,50){\makebox(0,0)[cc]{$\times$}}

\put(70,50){\makebox(0,0)[cc]{$\times$}}

\put(80,50){\makebox(0,0)[cc]{$\otimes$}}

\put(100,50){\makebox(0,0)[cc]{$\times$}}

\put(140,50){\makebox(0,0)[cc]{$\otimes$}}

\put(35,35){\makebox(0,0)[cc]{(a)}}

\put(75,35){\makebox(0,0)[cc]{(b)}}

\put(105,35){\makebox(0,0)[cc]{(c)}}

\put(135,35){\makebox(0,0)[cc]{(d)}}

\put(40,50){\makebox(0,0)[cc]{$\times$}}

\put(35,53){\makebox(0,0)[cc]{$q_1$}}

\put(45,58){\makebox(0,0)[cc]{$q_2$}}

\put(75,53){\makebox(0,0)[cc]{$q_1$}}

\put(105,58){\makebox(0,0)[cc]{$q_2$}}

\end{picture}

}

\def\fignine{

\def\JPicScale{0.8}
\ifx\JPicScale\undefined\def\JPicScale{1}\fi
\unitlength \JPicScale mm
\begin{picture}(125,90)(0,0)
\linethickness{0.3mm}
\put(30,30){\line(1,0){90}}
\linethickness{0.3mm}
\put(30,30){\line(0,1){60}}
\linethickness{0.3mm}
\put(30,70){\line(1,0){70}}
\linethickness{0.3mm}
\put(50,30){\line(0,1){10}}
\linethickness{0.3mm}
\put(30,40){\line(1,0){70}}
\linethickness{0.3mm}
\qbezier(35,70)(37.58,59.59)(41.19,52.38)
\qbezier(41.19,52.38)(44.8,45.16)(50,40)

\linethickness{0.3mm}
\put(100,30){\line(0,1){40}}

\put(35,35){\makebox(0,0)[cc]{(a)}}

\put(70,35){\makebox(0,0)[cc]{(b)}}

\put(35,50){\makebox(0,0)[cc]{(c)}}

\put(70,55){\makebox(0,0)[cc]{(d)}}

\put(130,30){\makebox(0,0)[cc]{$v\to$}}

\put(25,90){\makebox(0,0)[cc]{$x\uparrow$}}

\put(50,27){\makebox(0,0)[cc]{$v\simeq\alpha^{-2}$}}

\put(100,27){\makebox(0,0)[cc]{$v=e^{-2\pi t_c}$}}

\put(15,41){\makebox(0,0)[cc]{$x\simeq (2\pi\wt\lambda)^{-1}$}}

\put(22,70){\makebox(0,0)[cc]{$x= {1\over 4}$}}

\end{picture}

}

\def\figcooo{

\def\JPicScale{0.8}
\ifx\JPicScale\undefined\def\JPicScale{1}\fi
\unitlength \JPicScale mm
\begin{picture}(150,70)(0,0)
\linethickness{1mm}
\put(10,50){\line(1,0){15}}
\linethickness{0.2mm}
\put(25,50){\line(1,0){15}}
\linethickness{0.2mm}
\multiput(40,50)(0.12,-0.12){83}{\line(1,0){0.12}}
\linethickness{0.2mm}
\multiput(40,50)(0.12,0.12){83}{\line(1,0){0.12}}
\linethickness{0.2mm}
\multiput(50,60)(0.12,0.24){42}{\line(0,1){0.24}}
\linethickness{0.2mm}
\put(50,60){\line(1,0){10}}
\linethickness{1mm}
\put(70,50){\line(1,0){15}}
\linethickness{0.2mm}
\multiput(85,50)(0.12,0.12){83}{\line(1,0){0.12}}
\linethickness{0.2mm}
\multiput(85,50)(0.12,-0.12){83}{\line(1,0){0.12}}
\linethickness{0.2mm}
\multiput(95,60)(0.12,0.24){42}{\line(0,1){0.24}}
\linethickness{0.2mm}
\put(95,60){\line(1,0){10}}
\linethickness{1mm}
\put(120,50){\line(1,0){15}}
\linethickness{0.2mm}
\multiput(135,50)(0.12,0.12){83}{\line(1,0){0.12}}
\linethickness{0.2mm}
\put(135,50){\line(1,0){15}}
\linethickness{0.2mm}
\multiput(135,50)(0.12,-0.12){83}{\line(1,0){0.12}}
\put(25,30){\makebox(0,0)[cc]{(a)}}

\put(85,30){\makebox(0,0)[cc]{(b)}}

\put(135,30){\makebox(0,0)[cc]{(c)}}

\linethickness{1mm}
\put(160,50){\line(1,0){10}}
\linethickness{0.2mm}
\put(170,50){\line(1,0){10}}
\linethickness{0.2mm}
\multiput(180,50)(0.12,0.12){83}{\line(1,0){0.12}}
\linethickness{0.2mm}
\put(180,50){\line(1,0){10}}
\linethickness{0.2mm}
\multiput(180,50)(0.12,-0.12){83}{\line(1,0){0.12}}
\put(180,30){\makebox(0,0)[cc]{(d)}}

\put(32,53){\makebox(0,0)[cc]{$q_2$}}

\put(43,58){\makebox(0,0)[cc]{$q_1$}}

\put(177,53){\makebox(0,0)[cc]{$q_2$}}

\put(88,58){\makebox(0,0)[cc]{$q_1$}}

\end{picture}

}

\def\figtwo{

\def\JPicScale{0.5}
\ifx\JPicScale\undefined\def\JPicScale{1}\fi
\unitlength \JPicScale mm
\begin{picture}(100,70)(0,0)
\linethickness{0.7mm}
\multiput(20,70)(0.16,-0.12){250}{\line(1,0){0.16}}
\linethickness{0.7mm}
\multiput(60,40)(0.16,0.12){250}{\line(1,0){0.16}}
\linethickness{0.7mm}
\put(60,0){\line(0,1){40}}
\put(20,70){\makebox(0,0)[cc]{$\times$}}

\put(100,70){\makebox(0,0)[cc]{$\times$}}

\put(100,50){\makebox(0,0)[cc]{$\nearrow (0,\vec k-\vec \ell)$}}

\put(25,50){\makebox(0,0)[cc]{$(0,\vec \ell) \nwarrow$}}

\put(85,20){\makebox(0,0)[cc]{$\uparrow (k^0=0,\vec k)$}}

\end{picture}

}

\def\figredef{

\def\JPicScale{0.8}
\ifx\JPicScale\undefined\def\JPicScale{1}\fi
\unitlength \JPicScale mm
\begin{picture}(70,60)(0,0)
\linethickness{1mm}
\put(10,50){\line(1,0){20}}
\linethickness{0.2mm}
\put(30,50){\line(1,0){20}}
\linethickness{0.2mm}
\multiput(50,50)(0.24,0.12){83}{\line(1,0){0.24}}
\linethickness{0.2mm}
\multiput(50,50)(0.24,-0.12){83}{\line(1,0){0.24}}

\linethickness{1mm}
\put(90,50){\line(1,0){20}}
\linethickness{0.2mm}
\put(110,50){\line(1,0){20}}
\linethickness{0.2mm}
\put(110,35){\line(0,1){15}}

\put(30,50){\makebox(0,0)[cc]{$\times$}}

\put(50,50){\makebox(0,0)[cc]{$\times$}}

\put(110,50){\makebox(0,0)[cc]{$\times$}}

\put(40,30){\makebox(0,0)[cc]{(m)}}

\put(110,30){\makebox(0,0)[cc]{(n)}}

\end{picture}

}

\def\figtwob{

\def\JPicScale{0.5}
\ifx\JPicScale\undefined\def\JPicScale{1}\fi
\unitlength \JPicScale mm
\begin{picture}(100,70)(0,0)
\linethickness{0.7mm}
\put(60,0){\line(0,1){40}}
\put(60,40){\line(0,1){40}}

\put(60,40){\makebox(0,0)[cc]{$\times$}}

\put(60,80){\makebox(0,0)[cc]{$\times$}}

\put(82,60){\makebox(0,0)[cc]{$\uparrow (0,\vec k-\vec \ell)$}}

\put(85,20){\makebox(0,0)[cc]{$\uparrow (k^0=0,\vec k)$}}

\end{picture}

}

\def\figtwoc{

\def\JPicScale{0.5}
\ifx\JPicScale\undefined\def\JPicScale{1}\fi
\unitlength \JPicScale mm
\begin{picture}(100,70)(0,0)

\linethickness{0.7mm}
\put(60,0){\line(0,1){30}}
\put(60,50){\line(0,1){30}}

\linethickness{0.3mm}
\put(60,30){\line(0,1){20}}

\put(60,30){\makebox(0,0)[cc]{$\times$}}

\put(60,50){\makebox(0,0)[cc]{$\times$}}

\put(60,80){\makebox(0,0)[cc]{$\times$}}

\put(82,70){\makebox(0,0)[cc]{$\uparrow (0,\vec k-\vec \ell)$}}

\put(85,10){\makebox(0,0)[cc]{$\uparrow (k^0=0,\vec k)$}}

\end{picture}

}

\def\figtwod{

\def\JPicScale{0.5}
\ifx\JPicScale\undefined\def\JPicScale{1}\fi
\unitlength \JPicScale mm
\begin{picture}(100,70)(0,0)

\linethickness{0.7mm}
\put(60,0){\line(0,1){80}}

\put(60,80){\makebox(0,0)[cc]{$\otimes$}}

\put(85,10){\makebox(0,0)[cc]{$\uparrow (k^0=0,\vec k)$}}

\end{picture}

}

\begin{document}

\baselineskip 24pt

\begin{center}

{\Large \bf Scattering of D0-branes and Strings}

\end{center}

\vskip .6cm
\medskip

\vspace*{4.0ex}

\baselineskip=18pt

\begin{center}
\centerline{\large \rm Ashoke Sen$^1$ and Bogdan Stefa\'nski, jr.$^2$}

\vspace*{4.0ex}

$^1$ {\it International Centre for Theoretical Sciences - TIFR\\ 
Bengaluru - 560089, India}\\

$^3$ {\it Centre for Mathematics Innovation, City St. George's, University of London
\\ Northampton Square, EC1V 0HB London, UK}\\

\end{center}


\vspace*{1.0ex}
\centerline{\small E-mail:  ashoke.sen@icts.res.in, bogdan.stefanski.1@city.ac.uk}

\vspace*{5.0ex}

\centerline{\bf Abstract} \bigskip

It has been known for about thirty years that a scattering amplitude involving
D0-branes and closed strings suffers from infrared divergences beyond tree level.
These divergences arise because the conventional world-sheet approach 
cannot account for the difference between the D0-brane's momentum 
before and after scattering.
We show that, by
using string field theory, the divergence can be removed and the amplitude rendered finite and  unambiguous.
We illustrate this using the simplest possible example in bosonic string theory:  
a three-point
function with one incoming and one outgoing D0-brane  
and an incoming or outgoing
closed string tachyon.

\vfill \eject

\baselineskip=18pt

\tableofcontents

\sectiono{Introduction and summary} \label{sintro}

It has been known for many years that the scattering of D0-branes and closed
strings suffers from an infrared divergence\cite{9603156,9604014} at the next-to-leading order. 
The reason for this divergence
is also well understood and has been discussed in the original papers and subsequently\cite{9606102,9612064,9706115,
0106259,0406193,Evnin,0609216,1201.6606,1308.4016,2305.08297}. 
Physically, we expect that during such scattering the momentum
of the final D0-brane will differ from that of the initial D0-brane. However, in the
standard world-sheet approach the computation is done by summing over Riemann
surfaces with boundaries, with  D0-brane boundary conditions that have \textit{fixed} momenta. The
leading contribution comes from disk amplitudes and the next-to-leading order contribution
comes from the annulus amplitude. Hence, there is no scope of using different boundary
conditions corresponding to incoming and outgoing D0-branes.
It is a common phenomenon that use of incorrect external states leads to infrared
divergences and this case is no exception. One finds that the annulus amplitude
computed using fixed D0-brane boundary condition is infrared divergent\cite{9603156}.

String field theory (SFT) is well suited to address this problem.\footnote{This has been 
suspected by many people -- see in particular questions and comments by Emil
Martinec\cite{martinec} during Strings 2020 and by Igor Klebanov\cite{klebanov} 
during Strings 2025.} In this framework, the basic degrees of freedom are closed
string fields and 0+1 dimensional open string fields living on the D0-brane. Among
the open string fields are massless fields describing transverse displacement of the
D0-brane. Momentum carrying D0-branes can be regarded as coherent excitations
of these fields. Therefore, the problem of computing scattering amplitudes of
D0-branes and closed strings, with the incoming and outgoing D0-branes carrying
different momenta, can be reduced to the problem of computing the
matrix element of the interaction term involving closed and open string fields
between two different
coherent states of the open string field.

In this paper we illustrate this by computing 
the three point function in bosonic
string theory for which two of
the external states are D0-branes carrying different momenta 
and the third state is a
closed string tachyon. 
Generically three point functions of this kind vanish due to kinematic constraints,
but we can
avoid this by taking the external momenta to be complex. 
For example, if $M$ is the mass of the D0-brane and if we set $\alpha'=1$ 
so that the closed string tachyon has mass$^2$=$-4$, then
we can take  the incoming
D0-brane and closed string tachyon to have momenta 
\be\label{eq:in-momenta}
p_{\mbox{\scriptsize D0,in}}\equiv \left(E_{\mbox{\scriptsize D0,in}}\,,\,\vec p_{\mbox{\scriptsize D0,in}}\right)=(M, \vec k_1)\,,\qquad\qquad
p_{\mbox{\scriptsize tach,in}}
=(0, \vec k)\,,
\ee
and the outgoing D0-brane to have momenta 
\be\label{eq:out-momenta}
p_{\mbox{\scriptsize D0,out}}
=(M, \vec k_1)\,,
\ee
with 
\be
\vec k_1^2=0, \qquad \vec k^2=4, \qquad (\vec k+\vec k_1)^2=0\, .
\ee
String world-sheet theory offers a way
to calculate the
leading contribution to this amplitude: It is the disk one point function of the closed string
vertex operator $V$ with boundary conditions corresponding to a D0-brane at rest. 
Explicitly, this is given by\footnote{Using eq.(8.7.26) of \cite{polchinski} and eq.(4.123) of 
\cite{2405.19421},
 we get $M^2=2^{12}\, \pi^{23}/g_s^2$, but we shall not use this relation in our analysis.}
\be\label{e1.1n}
2\, \pi \, \delta(k^0) \times {1\over 2} \, g_s\, M\, ,
\ee
where $g_s$ is the string coupling, defined in the normalization convention for string
amplitudes given in \cite{2405.03784,2405.19421} 
and reviewed in appendix \ref{scollection}. 
The expression given in
\refb{e1.1n} calculates the T-matrix and the S-matrix element is obtained by
multiplying \refb{e1.1n} by $i$.
Note that \refb{e1.1n} ignores the fact
that the D0-brane states have spatial momenta $\vec k_1$ and
$\vec k_2\equiv \vec k_1+\vec k$. Since the D0-brane mass
$M$ is of order $1/g_s$ and a momentum $\vec k_i$ corresponds 
to a velocity $\vec v_i
=\vec k_i/M\sim g_s$, this discrepancy does not affect the leading order result. 
However, the absence of an explicit momentum conserving delta function captures
the fact that we are treating the D0-brane as a rigid classical object instead of
a quantum state.

Now consider the next order correction. Naively this will be given by
the annulus one point
function of the closed string tachyon vertex operator $V$. 
Let us parametrize the annulus
as a strip $0\le {\rm Re}\, w\le \pi$ in the complex $w$ plane with the
identification $w\equiv w+2\pi\, i\,  t$, and let 
the closed string vertex operator $V$ be inserted as a point $w_0$
with ${\rm Re}\, w_0 = 2\pi  x$.  
The standard rules of string theory require us to insert appropriate ghost
insertions that make the one point function into a volume form in the moduli 
space.
If $F(x,t)\, dx\, dt$ denotes this volume form, then
the amplitude
takes the form
\be \label{e1.1a}
\int_0^\infty dt\, \int_0^{1/4} dx\, F(x, t) \, ,
\ee
where we have restricted the integral over $x$
to be from 0 to 1/4 using the reflection symmetry
$x\to 1/2 - x$ on the annulus. 
Explicit computation gives
\ben\label{e5.14intro}
F(x,t) &=& 2\pi \delta(k^0)\,  { g_s\eta_c'\over \sqrt 2\pi}
\,  t^{-1/2} \, \eta(it)^{-24} 
\left[ {\vt_1(2x|it)\over \vt_1'(0|it)}\right]^{-2}\, , \qquad \eta_c'\equiv {1\over 2\pi}\, ,
\een
where
\be \label{e5.16aintro}
\vt_1(z|\tau) = - 2 \, e^{i\pi\tau/4}\, \sin(\pi z)\, \prod_{n=1}^\infty 
\{ (1- e^{2\pi i n\tau}) (1 - 2\, e^{2\pi i n \tau} \cos(2\pi z) + e^{4\pi i n\tau})\}\, ,
\ee
and
\be \label{e5.16bintro}
\eta(\tau) = e^{\pi i \tau/12} \prod_{n=1}^\infty \left(1- e^{2\pi i n\tau}\right)\, ,
\ee
are, respectively, the odd Jacobi theta function and Dedekind eta function.
We now
see that the integral over $F(x,t)$ has divergences from the 
$x\to 0$ and / or
$t\to \infty$ region. There are also divergences from the $t\to 0$ region associated with 
the closed-string tachyon of the bosonic string theory, which we will treat
using Witten's $i\epsilon$ prescription\cite{1307.5124}.

Our goal will be to use SFT to
extract an unambiguous, finite answer for \refb{e1.1a}. 
The main step in this analysis
is to represent the amplitude as a sum of
SFT Feynman diagrams and remove
the contribution due to the massless and tachyonic open string modes in the internal
propagators.
The contribution from the tachyon and massless ghosts can be treated using
the standard tools of SFT that was used, {\it e.g.}, for D-instantons.
Special treatment is needed for dealing with 
the collective modes that  describe the 
motion of the D0-brane
in the transverse directions. We  quantize them using standard tools of a
non-relativistic quantum mechanics that allows us to take the incoming and outgoing
D0-branes as momentum eigenstates. The matrix element of the SFT action interaction terms
between these states can then be used to compute the relevant part
of the effective action for external closed strings.
After computing the desired scattering amplitude using this effective action, 
we arrive at a finite, unambiguous  result for the amplitude. 

We shall now summarize our results. After using SFT to remove 
infrared divergences, the  
amplitude up to the first subleading order in the string coupling is given by
\be\label{efinalresult}
(2\, \pi)^{26}\, \delta^{(26)}(k_{\rm in} - k_{\rm out})\, \left[ \FF_{(0)} + \FF_{\rm annulus}
\right] \, .
\ee
Here $k_{\rm in}$ and $k_{\rm out}$ are the total incoming and
outgoing momenta and the incoming and outgoing D0-branes states are delta-function
normalized as in the case of a non-relativistic point particles. 
$\FF_{(0)}$ is the leading contribution from the disk amplitude:
\be
\FF_{(0)} = {1\over 2} \, g_s\, M\, ,
\ee
$M$ being the mass of the D0-brane.  $\FF_{\rm annulus}$ is the annulus 
contribution, given by
\be \label{eannulusF}
\FF_{\rm annulus} = \lim_{\alpha,\wt\lambda\to\infty\atop
\alpha/\wt\lambda = {\rm fixed}} \left[\FF_{(a)} +  \FF_{(b)} +  \FF_{(c)} +  \FF_{(d)} 
+  \FF_{(e)} +  \FF_{(f)} +  \FF_{(g)} + \FF_{\rm jac} +
\FF_{\rm cor}\right]\, ,
\ee
where,
\be
\FF_{(a)} = {1\over 4} \,  g_s\,  \eta_c'\,  \wt\lambda\,  \left\{-i + \erfi\left(\sqrt{2\, \ln\alpha}
\right)\right\}\, ,
\ee
erfi being the imaginary error function,
\be
\erfi(z) \equiv - {2\over \sqrt{\pi}} \, i\, \int_0^{iz} e^{-u^2} du
= {2\over \sqrt{\pi}} \, \int_0^{z} e^{u^2} du\, ,
\ee
\be
\FF_{(b)} =  -g_s\, \eta_c'\, {1\over 2\sqrt 2 } \left(1 +\alpha^{-2}\right) 
\, {\wt\lambda} \int_{t_c}^{{1\over 2\pi}
\ln(\alpha^2-1/2)} 
dt\,  t^{-1/2} \, \eta(it)^{-24}  \, ,
\ee
\be
\FF_{(c)}= -{1\over 2}\, g_s\, \eta_c' \,
\int_{1/(2\wt\lambda)}^1 {d\beta\over 4\, \beta^2} \, (1+\beta^2)\, 
\, \left\{-i+ \erfi\left(\sqrt{2 \ln\alpha + 2\ln{4\wt\lambda^2+1\over
4\wt\lambda} + 2 \ln {2\beta\over 1+\beta^2}}\right)\right\}
\, ,
\ee
\be
\FF_{(d)} =   { g_s\eta_c'\over \sqrt 2\pi} \int_A^{1/4} dx\,  \int_{t_c}^{B(x)} dt \, 
\,  t^{-1/2} \, \eta(it)^{-24} 
\left[ {\vt_1(2x|it)\over \vt_1'(0|it)}\right]^{-2}\, ,
\ee
\be \label{e5.6}
A \equiv (2\pi\wt\lambda)^{-1} (1-\alpha^{-2})\, ,
\ee
\be
e^{2\pi B(x)}  \equiv \alpha^2 \wt\lambda^2\sin^2(2\pi x) \, 
\left(1 + {1\over 4\wt\lambda^2}\right)^{2} 
\Bigg[1+ 2 \left\{ \cot^2(2\pi x)- \wt\lambda^2 f(\tan\pi x)^2\right\} \alpha^{-2}\wt\lambda^{-2}
\left(1 + {1\over 4\wt\lambda^2}\right)^{-2}\Bigg]\, ,
\ee
\be
\FF_{(e)} ={1\over 8} \, g_s\eta_c'\, \wt\lambda\, {1\over \sqrt{2\pi}} (\ln\alpha)^{-1/2}
 \, ,
\ee
\ben
\FF_{(f)} &=& { \eta_c' \, g_s\over \sqrt{2\pi}} \, \wt\lambda^2 \,
\int_{1/(2\wt\lambda)}^1 d\beta \, f(\beta)^2 \, 
{1\over 1+\beta^2}  {1\over  \sqrt{ \ln \alpha +
\ln{4\wt\lambda^2+1\over 4\wt\lambda}
+\ln{2\beta\over 1+\beta^2} }}
 \, ,
\een
\be
\FF_{(g)} = - {1\over 4} 
\, g_s\eta_c'\, \wt\lambda\, {1\over \sqrt{2\pi}} (\ln\alpha)^{-1/2} \, ,
\ee
\ben
\FF_{\rm jac} &=& {1\over \sqrt{2\pi}} \,
 \,  \Bigg[
-\ 2\,  g_s\, \eta_c'\,   \int_{1/(2\wt\lambda)}^1 d\beta 
\Bigg(\ln\alpha +\ln{4\wt\lambda^2+1\over 4\wt\lambda} +\ln{2\beta\over 1+\beta^2}
 \Bigg)^{-1/2} \nonumber \\ &&
\Bigg\{{25\over 8\beta} -{25\over 8}\beta+ \vec k^2 \, \tan^{-1}\beta\Bigg\}
\left({1\over \beta} 
- {2\beta\over 1+\beta^2}\right) \nonumber \\ &&
+\   \pi\, \vec k^2\, 
 g_s\, \eta_c'\,  \Bigg(\ln\alpha +\ln{4\wt\lambda^2+1\over 4\wt\lambda} 
 \Bigg)^{1/2}
\Bigg], \qquad \vec k^2 =4
\, ,
\een
and
\ben
\FF_{\rm cor}
&=& g_s \eta_c' \Bigg[ {1\over 4\sqrt{2\pi}} \wt\lambda^2 \alpha^{-2} \,
\int_{1/(2\wt\lambda)}^{1} {d\beta\over 1+\beta^2}\, \left\{\ln\alpha+\ln\wt\lambda + \ln{2\beta\over 1+\beta^2}\right\}^{-3/2} \, f(\beta)^4
\nonumber \\ &&
- {1\over 16\wt\lambda} \, \left\{- i + \erfi\left(\sqrt{2\ln\alpha}\right)\right\}
+{1\over 6\sqrt 2} \, \wt\lambda^{-1} \int_{t_c}^{{1\over 2\pi} \ln\alpha^2}
dt\, t^{-1/2} e^{2\pi t}\Bigg]\, .
\een
Here $f(\beta)$ is an arbitrary function of $\beta$ subject to the condition:
\be \label{eflimitintro}
f(1/2\wt\lambda)={4\wt\lambda^2-3\over 8\wt\lambda^2}, \qquad
 f(1)=0\, ,
 \ee
 and $\alpha$ and $\wt\lambda$ are parameters that need to be taken to be large but are
 otherwise arbitrary. More precisely, if we take $\alpha,\wt\lambda\sim\gamma$ for
 some large number $\gamma$, then the expression inside the square bracket in
 \refb{eannulusF} gives the correct expression for $\FF_{\rm annulus}$ up to
 correction terms of order $\gamma^{-1}$, possibly multiplied by powers of
 $\ln\gamma$. These corrections 
 represent contributions from massive open
 string modes in internal propagators but are suppressed for large $\alpha$, $\wt\lambda$.
 $\alpha$, $\wt\lambda$ and the function $f(\beta)$ arise in the formulation of SFT, but the final result is expected to be independent of these parameter since
SFTs corresponding to different choices of these parameters are
 related by field redefinitions. We have explicitly checked that the total contribution
to the term inside the square bracket in \refb{eannulusF} 
is independent of $\alpha$, $\gamma$ and $f(\beta)$ 
up to terms of order $\gamma^{-1}$.
 We can in principle avoid having to take the large $\alpha,\wt\lambda$ limit
 by adding appropriate terms of order
 $\gamma^{-1}$ to the expression inside the square bracket in \refb{eannulusF}
 that will make it fully independent of the choice of $\alpha$, $\wt\lambda$
 and $f(\beta)$, but we have not done this.
 
 The appearance of the imaginary error function can be traced to the open string
 tachyon propagating in the loop. We treat it using the usual $i\eps$ prescription
 and use the identity
 \be
 \int_{-i\infty}^{i\infty} {d\omega\over 2\pi} \,
 \kappa^{2+2\omega^2} {1\over \omega^2 
 +1 + i\eps} ={i\over 2} \left\{ -i + \erfi\left(\sqrt{2\ln \kappa}
 \right)\right\}, \qquad \hbox{for $\kappa>1$}\, .
 \ee
 
 Finally, we need to explain the lower cut-off $t_c$ on the integration over $t$. The integral
 is singular from the $t=0$ end due to the
closed string tachyon. These can also be dealt with using SFT\footnote{Note
that while the presence of the tachyon renders the theory inconsistent, {\it e.g.}
we do not have a unitary theory, there is no difficulty in getting finite
amplitudes involving internal tachyons using SFT. In the path integral,
this can be regraded as carrying out the integration over the tachyon field along its
steepest descent contour.} 
but we use Witten's
$i\eps$ prescription to deal with these singularities. This requires integrating $t$
up to some small number $1/\Lambda$, 
then change variable to $s=1/t$ and carry out the
integration over $s$ from $\Lambda$ to $\Lambda+i\infty$. So we can take
\be\label{eq:iepsilon-presc}
t_c = (\Lambda+i\infty)^{-1}\, .
\ee
The final result can be shown to be independent of $\Lambda$.

Numerical evaluation suggests the following result for $\FF_{\rm annulus}(i\vec k)$:
\be
\FF_{\rm annulus}(i\vec k) \approx (
7.28219 - 2.75650 \, i) g_s\, \eta_c' 
\approx (1.15900-0.43871\, i) g_s\,.
\ee
The imaginary part comes from closed string tachyons in intermediate states.

\sectiono{General strategy} \label{sproblem}

In this section we shall describe the origin of the divergence in the integral 
\refb{e1.1a} and the general strategy that we shall follow to resolve this, leaving the
details of the analysis to later sections.

\begin{figure}
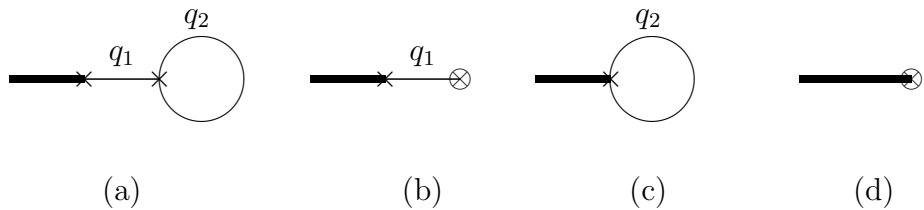

\begin{center}

\hbox{\figsix}

\vskip -1in

\caption{This figure shows the Feynman diagrams contributing to the annulus 
one point function of a closed string. The thick line represents external closed strings
and the thin lines denote internal open strings. The interaction vertex with a $\times$ represents part of a disk amplitude, while the interaction vertex with $\otimes$
represents part of an annulus amplitude. 
\label{figfive}
}
\end{center}
\end{figure}

The expression for the integrand $F(x,t)$ was given in \refb{e5.14intro} and the
actual derivation of this will be given in section \ref{sworld}. However,
one can
determine the general structure of the singularities of $F(x,t)$
even without explicit computation.
This is best understood using the language of SFT where the amplitude
is expressed as a sum over the Feynman diagrams of SFT of open and
closed strings and the divergences
appear from Schwinger parameter representation of the internal open string
propagators.  The relevant
Feynman diagrams were constructed in \cite{2012.11624,2210.11473} 
and are shown in Fig.~\ref{figfive}. In the full SFT there are also diagrams  with internal 
closed string propagators, but when the number of non-compact space-time 
dimensions is larger
than two, there are no divergences associated with the closed string propagators and we
can integrate out these modes and include their contribution as part of the interaction
vertex.  The exceptions are closed string tachyons whose effect will be discussed
separately in section \ref{sclosed}. 
In the Siegel gauge, an open string propagator is proportional to,
\be \label{el01}
L_0^{-1} = \int_0^1 dq \, q^{L_0-1}\, ,
\ee
and the divergences appear from the $q=0$ end of the integral due to 
$L_0\simeq 0$ or $L_0\le 0$ states. 
The relation between the variables $x,t$ and variables $q_1$
and $q_2$ associated with the two propagators shown in Fig.~\ref{figfive},  
were found in \cite{2012.11624} for
small $x$ and large $t$,
with the result:
\be\label{exqvq}
v\equiv e^{-2\pi t} \simeq q_2 / \alpha^2, \qquad x \simeq 
{q_1 / (2\pi\wt\lambda)}\, ,
\ee
where $\alpha$ and $\wt\lambda$ are two arbitrary large
parameters used in the construction
of SFT, with the final result expected to be independent of the choice of these
parameters. Therefore, the singularities of the Feynman diagram corresponding to small
$q_1$ will control the singularity of $F(x,t)$ for small $x$ and the singularities
corresponding to small $q_2$ correspond to the singularity of $F(x,t)$ for large $t$.
There are similar relations when only one of the variables $x$ and $v$ becomes small.
These will be discussed in section \ref{sbosonic}.

Note that \refb{el01} is valid for positive $L_0$. For $L_0\le 0$ the right hand side
of \refb{el01}
diverges, and the infrared divergences that we encountered earlier in \refb{e1.1a} 
from the $t=\infty$ and $x=0$ region all
stem from such contributions.  
The general strategy in SFT is to take the left hand sides of
\refb{el01} as the correct expression that should replace the divergent
integral on the right hand side of
\refb{el01}. This may still leave us with divergences from the $L_0=0$ states.
We need to identify the origin of these infrared divergences in SFT
and treat them correctly. 

Before going into the remedy, let us use \refb{exqvq} to
discuss what kind of divergence
we expect the integral in \refb{e1.1a} to possess. For this analysis we shall not
keep track of
the contribution from interaction vertices; hence we shall be able to extract the
leading divergent pieces only up to overall multiplicative constants.
The open string fields living on the D0-brane are (0+1) dimensional
fields. In momentum space they are functions of the energy variable $\omega$
which is manifestly conserved in amplitudes.
From Fig.~\ref{figfive} we see that
the open string propagator 1 is forced to carry zero energy since
the external closed string carries zero energy \refb{eq:in-momenta}. Furthermore, open strings on D0-branes
do not carry momenta along non-compact directions, so massless
open strings along this propagator have 
strictly vanishing $L_0$ and the integrand for small $x$ takes the form
\be \label{e1.4a}
dq_1 q_1^{-1} \simeq dx \, x^{-1}\, ,
\ee
where we used \refb{exqvq}. This gives $F(x,t) \sim x^{-1}$ for $x\to 0$ and the
integral diverges in this region. In bosonic string theory we also have open string
tachyon modes along propagator 1 
carrying $L_0=-1$. 
From \refb{el01} we see that these will produce divergent integrands of the form
\be \label{e1.5x}
dq_1\, q_1^{-2} \simeq {1\over 2\pi\wt\lambda}\,  dx \, x^{-2} \, .
\ee
The presence of the arbitrary constant $\wt\lambda$ may sound surprising, but this
is cancelled by the factors coming from interaction vertices. We shall see this explicitly
in section \ref{sbosonic}.

Let us now turn to propagator 2. It carries a loop energy $\omega$ 
that needs to be integrated. In the $\alpha'=1$ unit, 
massless open strings along this propagator have 
$L_0=-\omega^2$ and hence \refb{el01} leads to a divergent integrand
of the form\footnote{The 
large $t$ behaviour arises from the small $\omega$ region; so even though
there may be additional $\omega$ dependence from the vertex factors in the
graphs, they do not affect the leading contribution for large $t$.}
\be \label{e1.6a}
 dq_2\, \int_{-\infty}^\infty {d\omega\over 2\pi} \, q_2^{-\omega^2 - 1} = -2\, \pi\,  dt  \, 
 \int_{-\infty}^\infty {d\omega\over 2\pi}  \, \alpha^{-2\omega^2}\,
e^{2\pi t \omega^2} 
\simeq -{1\over \sqrt 2}\,  i\, dt \, t^{-1/2} \quad \hbox{for large $t$}\, ,
\ee
where we used $q_2 =\alpha^2 \, e^{-2\pi t}$ for large $t$ from \refb{exqvq}. 
The $i$ arises due to the fact that the integral needs to be defined by the euclidean
rotation $\omega \to i\omega_E$ of the integration contour.
This shows $F(x,t)\sim t^{-1/2}$ for
large $t$.  Hence, the integral diverges in this region, as was originally noted in\cite{9603156,9604014}.
On the other hand
a tachyon appearing in propagator 2 will have $L_0=-\omega^2-1$, and
integration over $\omega$ will lead to
a contribution of the form
\be\label{etach2}
dq_2\, \int_{-\infty}^\infty {d\omega\over 2\pi} \, q_2^{-\omega^2 - 2} = -2\, \pi\,
dt  \, \int_{-\infty}^\infty {d\omega\over 2\pi} \,  
\alpha^{-2-2\omega^2}
e^{2\pi t(\omega^2+1)} \simeq -{1\over \sqrt 2}\,   i\,  \alpha^{-2}\, dt\,
t^{-1/2} \, e^{2\pi t} \quad \hbox{for large $t$}\, .
\ee
Thus the $t$ integral diverges from the large $t$ region. The singularities appearing
in \refb{e1.4a} - \refb{etach2} 
match the singularities of $F(x,t)$ given in \refb{e5.14intro} in the small $x$ and
large $t$ limit.

To remedy these problems, we need to first identify the massless and tachyonic
open string fields
that are responsible for this divergence. 
In bosonic open string theory the massless string
 fields can be represented as
 \be
 |\Psi\rangle = |\Psi_s\rangle + |\Psi_{ns}\rangle\, ,
 \ee
where the Siegel gauge field $|\Psi_s\rangle$ takes the form\footnote{Since we shall
work in a flat background metric, all our spatial indices will be raised and lowered
by the flat metric $\delta_{ij}$ and we shall not distinguish between upper and lower
indices.}
\ben \label{e1.8cc}
 |\Psi_s\rangle &=& \int {d\omega\over 2\pi}\, 
 \bigg[y^i(\omega) \alpha^i_{-1}c_1 |\omega\rangle
+ a_0(\omega)  \alpha^0_{-1} c_1 |\omega\rangle 
+ i\,  q(\omega)  |\omega\rangle +i\,  p(\omega) c_{-1}
c_1 |\omega\rangle
\bigg], \nonumber \\
&&\hskip 1in   |\omega\rangle \equiv e^{-i\omega X^0}(0) |0\rangle,
\qquad  \alpha^\mu_{-1}|\omega\rangle \equiv i\sqrt 2 \p X^\mu
e^{-i\omega X^0}(0) |0\rangle \, ,
\een
and the out of Siegel gauge field $|\Psi_{ns}\rangle$ takes the form:
\be \label{e1.8bc}
|\Psi_{ns}\rangle = \int {d\omega\over 2\pi}\, \bigg[\tilde y^i(\omega) \alpha^i_{-1}
c_0 c_1 |\omega\rangle
+ \tilde a_0(\omega) \, \alpha^0_{-1} \, c_0 \,
c_1 |\omega\rangle 
+i\, \tilde q(\omega) \,  c_{-1} \, c_0\, 
c_1 |\omega\rangle 
+ i\, \tilde p(\omega) 
c_0  |\omega\rangle
\bigg] \, .
\ee
Here $|\omega\rangle$ represents the Fock vacuum carrying 
energy $\omega$, $\alpha^\mu_n$ are oscillators associated
with world-sheet scalar fields $X^\mu$, and $b,c$ are
world-sheet diffeomorphism ghosts. 
$y^i$ represent the zero modes 
associated with the transverse
position of the D0-brane along the flat directions, $a_0$ is the U(1) gauge field 
living on the D0-brane world-volume and $p$, $q$ are the ghost fields associated with
gauge fixing of the U(1) gauge symmetry on the D0-brane world-volume. 
While propagating along the propagators 1 or 2, these fields will be responsible for
the divergences in the $x\to 0$ and $t\to\infty$ limit. 
The fields $\tilde y^i$, $\tilde p$, $\tilde  a_0$ and $\tilde q$ are the anti-fields of $y^i$,
$p$, $a_0$ and $q$ respectively. In Siegel gauge $|\Psi_{ns}\rangle$ is set to
zero. Thus we see that the gauge fixing follows the
usual Batalin-Vilkovisky (BV) 
formalism where we integrate over half of the fields, setting their
anti-fields to zero.

We shall use the normalization 
conventions given in \cite{2405.19421}, in which
\be \label{enormfo2}
\langle \omega|c_{-1} c_0 c_1|\omega'\rangle=-2\pi K\delta(\omega
+\omega')\, ,
\ee
where
$\langle \cdots \rangle$ is the 
disk correlation function, and $K$ is a constant, related to the D0-brane mass
$M$ and the closed string coupling $g_s$ by the relation\cite{2405.19421}:
\be\label{e1.12aa}
 K= 
- g_s \, {M\over 2\sqrt {\eta_c}}, \qquad \eta_c\equiv {i\over 2\pi}\, .
\ee
For all the world-sheet fields we use the normalization given in \cite{2405.19421}.
In computing correlation functions on the upper half plane, we shall often
use the doubling trick to express the correlation function in the full complex plane
via the replacement,
\ben
&& \bar b(x+iy) \to b(x-iy), \quad  \bar c(x+iy) \to c(x-iy),  \quad
\bar \p X^\mu (x+iy)
\to \pm \p X^\mu (x-iy), \nonumber \\
&& e^{ik.X}(x+iy) = e^{ik.(X_L+X_R)}(x+iy) \to e^{i k.X_R}(x+iy)\, 
e^{\pm i k. X_R}(x-iy)\, ,
\een
where we choose $+$ sign for coordinates with Neumann boundary condition and
$-$ sign for coordinates with Dirichlet boundary condition. Here $X_R$ and $X_L$
denote the holomorphic and anti-holomorphic components of $X$. In this spirit, the
open string vertex operator $e^{i\omega X^0}(x)$, inserted on the real
line  in the upper half plane, may be
regarded as $e^{2i\omega X_R^0}(x)$ in the full complex plane.

The action $S$ will be normalized such that $e^S$ is the weight factor in the
Euclidean path integral. 
In this convention, the
kinetic term of the string field in the Siegel gauge
gives\footnote{In this computation we need to account for the fact that when the
Grassmann odd fields $p$, $q$ pass through the BRST charge $Q_B$, we get an
extra minus sign.}
 \be \label{e1.8a}
 {1\over 2}\, \langle \Psi_s| Q_B|\Psi_s\rangle = K\, \int {d\omega\over 2\pi} \, \bigg[
 {1\over 2} \omega^2 y^i(-\omega) y^i(\omega) - {1\over 2} \omega^2 
 a_0(-\omega) a_0(\omega) +
 \omega^2 p(\omega)
 q(-\omega)
 \bigg]\, .
 \ee
 The $\omega^2$ factors in the quadratic terms are responsible for the $\omega^{-2}$
 terms in the propagator, which in turn produce the divergences in the $x\to 0$ and / or
 $t\to\infty$ limit via \refb{e1.4a} and \refb{e1.6a}. Indeed, not only the divergent part but
 the complete world-sheet expression for the amplitudes can be shown to emerge from
 the Feynman rules in Siegel gauge.
 
 We now describe how to treat 
 the divergences coming from the $y^i$, $a_0$, $p$ and $q$ fields.
 We begin with the $p,q$ fields. 
  In the Faddeev-Popov formalism,
  these arise from fixing the gauge transformation
 associated with the parameter
 \be \label{eLambda}
 |\Lambda\rangle =  i\, \int {d\omega\over 2\pi} \, \theta(\omega) \, |\omega\rangle + \cdots\, ,
\ee
by setting to zero the out of Siegel gauge field $\tilde p$  
appearing in \refb{e1.8bc}.
Indeed, 
using the gauge transformation law $\delta |\Psi\rangle = Q_B|\Lambda\rangle$, we get
\be
\delta \tilde p = -\omega^2 \, \theta\, .
\ee
Therefore, fixing the $\tilde p=0$ gauge produces a jacobian factor $\omega^2$ which is
captured by the integration over the ghost fields $p$, $q$, and in turn leads to the
infrared divergence problems discussed earlier. The remedy we shall follow is to fix
a different gauge, setting 
\be
a_0=0\, .
\ee
This has two effects. First of all it removes integration over
$p$ and $a_0$. The second effect is that we now have to integrate over
the out of Siegel gauge 
field $\tilde p$ and a new  ghost field which we can identify as the anti-field
$\tilde  a_0$ of the field $a_0$, appearing in \refb{e1.8bc}. 
The string field at level 0 in this new gauge now takes the form
\be \label{epsinew}
|\Psi_{new}\rangle = \int {d\omega\over 2\pi}\, \bigg[y^i(\omega) \alpha^i_{-1}c_1 |\omega\rangle
+i\, q(\omega) \,  |\omega\rangle   + \tilde  a_0(\omega) \, \alpha^0_{-1} \, c_0  \, 
c_1 |\omega\rangle + i\, \tilde p(\omega) 
c_0 |\omega\rangle 
\bigg] \, .
\ee
It will also be important to determine the integration measure over these fields. For this
we note that from the perspective of the BV formalism,
the part of the level zero string field that has been fixed to zero in this gauge
is
\be \label{epsigf}
|\Psi_{gf}\rangle=
\int {d\omega\over 2\pi}\, \bigg[\tilde y^i(\omega) \alpha^i_{-1} c_0 c_1 |\omega\rangle
+i\, \tilde q(\omega) \,  c_{-1} \, c_0\, 
c_1 |\omega\rangle 
+ a_0(\omega)  \alpha^0_{-1}c_1 |\omega\rangle 
+ i\, p(\omega) c_{-1} c_1  |\omega\rangle\bigg] \, .
\ee
In this expansion, $\tilde y^i$, $p$, $a_0$ and $\tilde q$ are the anti-fields of
$y^i$, $\tilde p$, $\tilde a_0$ and $q$ respectively up to constant normalizations and
signs. Following the standard rules of BV quantization,
the gauge invariant measure can now
be written as
\be 
\int Dy^i\,  D\tilde y^i \, D \tilde p\, D p\, D\tilde a_0 \, Da^0\, Dq\, D\tilde q
\prod_i \delta(\tilde y^i) \delta (p) \, \delta (a^0) \, \delta(\tilde q)
= \int Dy^i\,  D \tilde p\, D\tilde a_0 \,  Dq\, ,
\ee
without any additional Jacobian factor.

In this gauge the action  contains the terms
\be \label{e1.17a}
 {1\over 2}\, \langle \Psi_{new}| Q_B|\Psi_{new}\rangle
 = K \int {d\omega\over 2\pi}  \bigg[  {1\over 2} \omega^2 y^i(-\omega) y^i(\omega)
- \tilde p(-\omega)\tilde p(\omega) +  {i}\,  \sqrt 2\, 
\,  \omega \, \tilde a_0(-\omega) \, q(\omega)
\bigg]\, ,
\ee
showing that $\tilde p$ plays the role of an auxiliary field and the kinetic term of the new
Faddeev-Popov ghosts $q$ and $\tilde a_0$ is proportional to $\omega$. The latter is
related to the fact that for the gauge transformation 
parameter $|\Lambda\rangle$ given in
\refb{eLambda},  $\delta |\Psi\rangle=Q_B|\Lambda\rangle$ gives 
$\delta a_0\propto i\, \omega\, \theta$. The path integral
over the $\tilde p$ field now gives a finite result. On the other hand the $\tilde a_0$-$q$ ghost 
propagator is proportional to $\omega/(\omega^2+i\eps)$ 
and hence the potentially divergent integrand
involving the $\tilde a_0$-$q$ propagator
has the form $d\omega\, f(\omega) \, \omega  / (\omega^2 +i\eps) $ for some smooth function $f(\omega)$. The integral over $\omega$ now 
has no divergence from the $\omega\simeq 0$ region.

This leaves us to deal with the modes labelled by $y^i$. 
Physically the origin of the divergence is clear. In the approach discussed
so far, the D0-brane is taken to be a static object that does not backreact during the
scattering process. In particular, the D0-brane boundary condition breaks
translation invariance along the non-compact spatial directions and hence violates
conservation of spatial momenta. This is reflected in the fact that in the standard
world-sheet approach both the initial and the
final D0-brane states are taken to be zero momentum objects. 
In actual practice, the D0-brane will suffer a recoil, leading to momentum conservation.
The choice of `wrong
external states' typically leads to infrared divergences in the amplitudes and this is
precisely what is responsible for the divergence from the $y^i$ propagators. 

To remedy this problem, we note that the 
$y^i$'s are the collective modes
of the D0-brane associated with the motion in the transverse direction. 
As is well known, the
collective modes of solitons cannot be treated using perturbation theory, 
instead they
have to be quantized separately. So we need to first
remove the contribution of the 
$y^i$'s
from the open string propagator. This removes the remaining
terms responsible for the
divergences and allows us to integrate over the other open string modes to
construct an effective theory of the collective modes. We then quantize the
collective modes. 
This  will 
generate the momentum eigenstates of the D0-brane that appear in the initial and
final states of a scattering process and will allow us to have initial and final D0-branes
carrying different momenta, leading to overall momentum conservation. 
This will be discussed in section \ref{stoy}. 

A further complication arises due to the fact that the $y^i$'s that enter the expansion
of the string field are not directly the collective coordinates, but are related to 
them by field redefinition.
Since our starting point is the SFT action, we need to find this field
redefinition and then express the SFT action in terms of the collective
modes, including the effect of the Jacobian factor due to the change of variables
in the path integral.
In section \ref{scollective} we shall find this field 
redefinition by comparing the coupling of $y^i$
to a set of closed string states to the expected coupling of the collective modes to the
closed string states.

In bosonic string theory we also have 
integration over the open string tachyon field $T(\omega)$
that appears in the expansion of the string field as
\be
|\Psi\rangle =\int {d\omega\over 2\pi} \, T(\omega) \, c_1|\omega\rangle\, .
\ee
The action involving the tachyon field takes the form:
\be
{K\over 2}
\int_{-\infty}^\infty \, {d\omega\over 2\pi} \, (1+\omega^2) \, T(-\omega) T(\omega)\, ,
\ee
leading to a tachyon propagator $-(\omega^2+1)^{-1}$. This can be traced to the
fact that $c_1|\omega\rangle$ has $L_0$ eigenvalue $-(\omega^2+1)$ and
is responsible for the leading divergences appearing in \refb{e1.5x} and
\refb{etach2} via \refb{el01}. Let us begin with \refb{e1.5x} that comes from a tachyon 
propagator with $\omega=0$. 
The remedy of this is to replace the right hand side of \refb{el01} by the
left hand side of \refb{el01} with $L_0=-1$,
i.e.\ after representing
the contribution from the $x=0$ region in terms of the variable $q_1$ coming from SFT, we simply replace $\int_0^1 dq_1\, q_1^{-2}$ by $-1$.

For \refb{etach2}, i.e.\
when the tachyon appears in propagator 2, the situation is a bit more complicated.
As before, we replace $\int_0^1 dq_2 q_2^{-\omega^2-2}$ by $(\omega^2+1)^{-1}$.
Although the integral over $\omega$
looks divergence free, in SFT the interaction vertices will typically contain
terms proportional to $\exp(C\omega^2)$ for some positive constant $C$. Hence, the
end points of the $\omega$ integration contour must approach $\pm i\infty$ in order
that the integral converges\cite{1604.01783}. If we take
the integration contour to be along the imaginary 
$\omega$ axis then the integrand develops a
pole on the integration contour at $\omega=\pm i$. Hence we need to choose an 
integration contour avoiding these poles. We shall use the standard $i\eps$ prescription
where we replace $p^2+m^2$ in the denominator by $p^2+m^2-i\eps$. This corresponds
to replacing $(\omega^2+1)^{-1}$ by $(\omega^2+1+i\eps)^{-1}$. This ensures that
we do not encounter any poles during the Wick rotation from the real $\omega$ axis to the
imaginary $\omega$ axis. For definiteness, we shall work with this choice of
contour.
However, given that
the tachyonic mode represents an instability of the D0-brane and the system is
not physical, other choice of contour may also be possible.

\sectiono{Bosonic collective modes} \label{stoy}

Our goal in this section will be to deal with the divergences associated with the $y^i$
propagators. 
For this we first integrate out all the open string fields other than the $y^i$'s
to construct an effective action of $y^i$ and the closed string tachyon field $\Sigma$
whose amplitude we are trying to compute. 
The analysis of the previous section shows that this 
process does not encounter any infrared divergence.
If we try to compute amplitudes from this effective action using perturbation
theory, we shall encounter infrared divergences due to the $\omega^{-2}$
singularity in the $y^i$ propagators. Therefore we cannot treat the $y^i$'s using
perturbation theory.
We shall now describe how this difficulty is resolved.

If $y^i$'s had been exactly the collective modes, then  we
could remove the contribution of the $y^i$ propagators from the
internal lines of an open string propagator and quantize the $y^i$'s 
separately to construct momentum 
eigenstates of the D0-brane, which can then be used for computing a scattering amplitude. The complication arises from the fact that the $y^i$'s are not exactly the
collective modes. For example, the action should be invariant under a rigid
translation of a collective mode associated with broken translation invariance, but
the full SFT action is not invariant under such translations of
$y^i$.  
Let $\chi^i(t)$ be the actual collective mode, 
related to $y^i(t)$ and
other string field components by a field redefinition. 
As will be discussed in section \ref{scollective}, 
we can find this field redefinition using perturbation theory,
and the Jacobian due to the change of variable from $y^i$ to $\chi^i$ will give an
additional term in the effective action. This resulting
action should be invariant under the transformation\footnote{The action should also
have Lorentz invariance. However, since the D0-brane has mass of order $g_s^{-1}$,
the Lorentz transformation will mix different orders in $g_s$ expansion. For this reason
we do not make use of Lorentz symmetry.}
\be \label{e2.1c}
\Sigma (t,\vec x) \to \Sigma (t,\vec x+\vec a),
\qquad \chi^i(t)\to \chi^i(t)-a^i\, , 
\ee
for any constant vector $\vec a$. 
Also,  the action
is invariant under a time reversal symmetry
\be\label{etime}
\Sigma (t,\vec x) \to \Sigma (-t,\vec x),
\qquad \chi^i(t)\to \chi^i(-t)\, ,
\ee
and parity symmetry
\be\label{eparity}
\Sigma (t,\vec x) \to \Sigma (t,-\vec x),
\qquad \chi^i(t)\to -\chi^i(t)\, .
\ee
An action of this type is
\ben\label{e2.2c}
&-& {1\over 2} \int dt\, d^{D}x\,  \left[\eta^{\mu\nu}\, 
\p_\mu\Sigma (t,\vec x)\,  \p_\nu \Sigma  (t, \vec x) + m_\Sigma^2 
\Sigma  (t, \vec x)^2\right] +
{M\over 2} \int dt \, \p_t \vec\chi(t) . \p_t \vec\chi(t) 
 \nonumber \\ &
+& \int dt \,  \FF(\vec\nabla)  \, \Sigma (t, \vec \chi(t)) +
\int dt \,  \p_t \chi^i(t) \, \p_t\chi^j(t)\, \FF(\vec\nabla)_{  ij} \, \Sigma (t, \vec \chi(t))
+\cdots\, ,
\een
where we have kept terms  up to linear order in $\Sigma$ and quadratic order in
$\chi^i$ (other than those appearing in the argument of $\Sigma$) since these are
the terms relevant for the Feynman diagrams of Fig.~\ref{figfive}.
Here $m_\Sigma$ is the mass of $\Sigma$,
$M\propto g_s^{-1}$ is the mass of the D0-brane and $\FF(\vec\nabla) , \,
 \FF(\vec\nabla)_{  ij}$
are polynomials of spatial derivative operators acting on $\Sigma $. 
$D$ is the number of non-compact spatial dimensions which will eventually be set to
25.
$\FF(\vec\nabla)  \, \Sigma (t, \vec \chi(t))$ means that we first compute 
$\FF(\vec\nabla)  \, \Sigma (t, \vec x)$ and then replace $\vec x$ by $\vec \chi$.
These spatial derivatives will translate into
factors of spatial momenta in the final amplitude. On the other hand, since the
external closed string carries zero energy, the time derivative of $\Sigma$ can be
ignored. Also, we have used the time reversal symmetry to exclude terms
linear in $\p_t\chi^i$ and ignored terms proportional to $\p_t^n\chi^i$ for $n\ge 2$ since
they can be removed by field redefinition.
The full theory contains many  more terms  consistent 
with the symmetry \refb{e2.1c} involving higher powers of $\p_t\chi^i$. However,
these terms will not contribute to the Feynman diagrams appearing in 
Fig.~\ref{figfive} and hence will not be relevant for our discussion. They will of course
become important at higher order in $g_s$ expansion.

We shall now show that the term proportional to $\FF(\vec\nabla)_{  ij}$ also does not affect
our analysis. For this we make a field redefinition
\be\label{efieldredef}
\chi^j \to \chi^j- M^{-1} \, \chi^i(t) \,  \FF(\vec\nabla)_{ij} \, \Sigma (t, \vec \chi(t))\, .
\ee
This removes the terms in \refb{e2.2c} proportional to $\FF(\vec\nabla)_{ij}$ 
and produces
new terms involving $\p_t\Sigma$ or cubic terms in $\chi^i$. Since the external
closed string has zero energy and since the Feynman diagrams in Fig.~\ref{figfive}
involve interaction terms that are at most quadratic in the $y^i$'s, 
none of these new terms contribute to the Feynman
diagrams of Fig.~\ref{figfive}. The field redefinition \refb{efieldredef}
will give rise to a Jacobian whose effect will be to give 
an additional term in
the effective action proportional to $\int dt  \FF(\vec\nabla)_{  ii} \Sigma (t, \vec \chi(t))$,
but this just renormalizes $\FF(\vec\nabla) $. 
Thus we work with the action
\ben \label{e3.8xx}
&-& {1\over 2} \int dt\, d^{D}x\,  \left[\eta^{\mu\nu}\, 
\p_\mu\Sigma (t,\vec x)\,  \p_\nu \Sigma  (t, \vec x) + m_\Sigma^2 
\Sigma  (t, \vec x)^2\right] +
{M\over 2} \int dt \, \p_t \vec\chi(t) . \p_t \vec\chi(t) 
 \nonumber \\ &
+& \int dt \,  \FF(\vec\nabla)  \, \Sigma (t, \vec \chi(t)) \, .
\een

It is instructive to see what happens if we expand the action in powers of
$\chi^i$ and evaluate the 
amplitude by computing Feynman
diagrams. The action to quadratic order in $\chi^i$ takes the form:
\ben \label{e2.7c}
&&\hskip -.3in  -{1\over 2} \int dt\, d^{D}x\,  \left[\eta^{\mu\nu}\, 
\p_\mu\Sigma (t,\vec x)\,  \p_\nu \Sigma  (t, \vec x) + m_\Sigma^2 
\Sigma  (t, \vec x)^2\right]
+ {M\over 2} \int dt \, \p_t \vec\chi(t) . \p_t \vec\chi(t) 
\nonumber \\ && \hskip -.3in +  
\int dt \,  \left(\FF(\vec\nabla)  \,\Sigma (t, \vec 0) + \chi^i(t)\, \FF(\vec\nabla)   \p_i
\Sigma (t, \vec 0) 
+ {1\over 2} \chi^i(t) \chi^j(t) \, \FF(\vec\nabla)  \p_i \p_j \Sigma (t, \vec 0)  \right)
+ \cdots
\, ,
\een
where $\p_{i_1}\cdots \p_{i_n}\Sigma(t,\vec 0) \equiv
\p_{i_1}\cdots \p_{i_n}\Sigma(t,\vec x) |_{\vec x=\vec 0}$.
In Fourier transformed space, the $\chi^i$ propagator will be of order $\omega^{-2}$.
Hence loops of the $\chi^i$ field will lead to infrared divergences, {\it e.g.} in
diagrams in Fig.~\ref{figfive}(c) when the propagator 2 is a $\chi^i$ field and 
the interaction vertex is produced by the last term in \refb{e2.7c}.

We now describe the remedy of the problem, which is to quantize the mode
$\chi^i$ exactly instead of treating it using Feynman diagrams.
While we can use the path integral formulation for both the
collective coordinate $\chi^i$ 
and the field $\Sigma $, it will be a bit more illuminating to treat the
modes $\chi^i$ using the Hamiltonian formulation. More precisely we use the Routhian
formalism where we use the Hamiltonian formalism for
the fields $\chi^i(t)$ and the  Lagrangian formalism for the
field $\Sigma$.
We define the conjugate momenta
\be
p_j= {\p L\over \p (\p_t \chi^j)} = M\, \p_t\chi^j \, ,
\ee
and the Routhian:
\ben
R = \sum_j p_j \p_t \chi^j - L
&=& {p_j p_j \over 2M} + {1\over 2} \int dt\, d^{D}x\, \left[ \eta^{\mu\nu}\, 
\p_\mu\Sigma (t,\vec x)\,  \p_\nu \Sigma  (t, \vec x)  +{1\over 2} m_\Sigma^2 
\Sigma(t,\vec x)^2\right] \nonumber \\ &&
-\ \int dt \,  \FF(\vec\nabla)  \,\Sigma (t, \vec \chi(t))  \, .
\een
We shall treat $p_ip_i/(2M)$ as the unperturbed Hamiltonian of the collective mode 
and 
take the incoming and the outgoing D0-brane to be eigenstates of energy
and momentum $(\omega_{in}, \vec k_{in})$ and
$(\omega_{out}, \vec k_{out})$ respectively, satisfying
\be
\omega_{in} = M+{\vec k_{in}^2\over 2M}, \qquad
\omega_{out} = M+{\vec k_{out}^2\over 2M}\, .
\ee
Our strategy will be to first compute the matrix element of
$\FF(\vec\nabla)  \,\Sigma (t, \vec \chi(t))$ between
the incoming and outgoing D0-brane states $|\vec k_{in}\rangle$, 
$|\vec k_{out}\rangle$, and then treat 
$\langle \vec k_{out}| \int dt\, \FF(\vec\nabla)  \,\Sigma (t, \vec \chi(t))
|\vec k_{in}\rangle$ as a term in the effective action of $\Sigma$ from
which we can compute the one point function of $\Sigma$. This will give the
desired amplitude.

We normalize the D0-brane states as those of a non-relativistic particle so as not to
mix different orders in perturbation theory
\be
 \langle \vec k_{out}|  \vec k_{in}\rangle = 
(2\pi)^D \delta^{(D)} (\vec k_{in}-\vec k_{out})\, ,
\ee 
and introduce the Fourier transform $\wt\Sigma$ of $\Sigma$, defined through
\be
\Sigma(t,\vec x) = \int {d\omega\over 2\pi} \int {d^D k\over (2\pi)^D} \, 
e^{-i\omega t + i \vec k.\vec x} \, \wt\Sigma(\omega, \vec k)\, .
\ee
This gives the linear term in the effective action of $\Sigma$ to be
\ben
&& 
\int dt\, \langle \vec k_{out}|  \FF(\vec\nabla)  \,\Sigma (t, \vec \chi(t))
| \vec k_{in}\rangle
= \int dt\,  e^{i(\omega_{out}-\omega_{in})t} \langle
\vec k_{out}|  \FF(\vec\nabla)  \,\Sigma (t, \vec \chi(0))
|\vec k_{in}\rangle \nonumber \\
&=& \int dt\, \int {d\omega\over 2\pi} \int {d^Dk\over (2\pi)^D} \, e^{i(\omega_{out}-
\omega_{in}-\omega)t}\,   \FF(i\vec k) \, 
\wt\Sigma(\omega, \vec k)
\langle
\vec k_{out}|  e^{i\vec k.\vec \chi(t=0)}
|\vec k_{in}\rangle \nonumber \\
&=&  \int {d\omega\over 2\pi} \, \int {d^Dk\over (2\pi)^D} \, \FF(i\vec k) \, 
\wt\Sigma(\omega, \vec k)\, 2\pi \, \delta (\omega_{out}-\omega_{in}-\omega)\, 
(2\pi)^D \delta^{(D)} (\vec k+\vec k_{in}-\vec k_{out})\, . \nonumber \\
\een
Above, the first delta function comes from the $t$ integral and the second delta
function comes from the matrix element. 
The relevant amplitude for an external closed string state
of energy $\omega$ and momentum $\vec k$ is then given by,
\be \label{efinresult}
 \FF(i\vec k) \, 
 2\pi \, \delta (\omega -\omega_{out}+\omega_{in})\, 
(2\pi)^D \delta^{(D)} (\vec k+\vec k_{in}-\vec k_{out})\, .
\ee
This recovers the full energy-momentum conserving delta function. 
Thus the main task is to compute $\FF(i\vec k)$ using SFT.

\sectiono{Field redefinition from the string fields to collective coordinates} 
\label{scollective}

In this section we shall describe the procedure for finding the field redefinition that
takes us from string field variables $y^i$ to the collective coordinates $\chi^i$. 

To simplify notation, 
for the open string field $y^i$ and the collective coordinate $\chi^i$ we shall use
the same symbol for the field and its Fourier transform in the time variable -- the
argument of the field will convey information about which one we are using.
We shall be looking for a relation between $y^i$ and $\chi^i$ of the form
\ben\label{e5.2gen}
y^i(\omega) &=& \sqrt{M\over K} \chi^i(\omega) + 
\int {d^D k\over (2\pi)^D}\, g^i(\omega, \vec k)\wt\Sigma(\omega, \vec k)
\nonumber \\ &+&
\int {d\omega'\over 2\pi} 
\int {d^D k\over (2\pi)^D}\, f^i{}_{j}(\omega, \omega', \vec k)\,  \chi^j(\omega') \, 
\wt\Sigma(\omega-\omega', \vec k) +\cdots\, ,
\een
where $\cdots$ denotes other terms containing higher powers of $\chi^i$ and / or
$\wt\Sigma$ that will not be needed in our analysis
and $f^i{}_{j}$ and $g^i$ are functions that we need to determine. 
The normalization factor $\sqrt{M/K}$ in the first term has been fixed by comparing the
kinetic term of $y^i$ in the SFT action \refb{e1.8a} 
and that of $\chi^i$ in the
action of collective field theory given in \refb{e2.2c}.
Rotational invariance prevents the appearance of terms of the form
$\int d\omega' C^i{}_{jk}(\omega,\omega') \chi^j(\omega')\chi^k (\omega-\omega')$ on the
right hand side of \refb{e5.2gen}.
Treating $\wt\Sigma$ as a background field, we can write,
up to an overall constant normalization,
\ben\label{e4.2xx}
&& \prod_{i,\omega} dy^i(\omega) \simeq \prod_{i,\omega} d\chi^i(\omega) 
\left[1 + \sqrt{K\over M} \int_{-i\infty}^{i\infty} (-i) {d\omega'\over 2\pi} 
\int {d^D k\over (2\pi)^D} \, f^{j}{}_{j}(\omega', \omega', \vec k) 
\,  \wt\Sigma(0, \vec k)\right]
\nonumber \\
&\simeq&
\prod_{i,\omega} d\chi^i(\omega) \exp\left[-i \sqrt{K\over M} 
\int_{-i\infty}^{i\infty}{d\omega'\over 2\pi} 
\int {d^D k\over (2\pi)^D} \, f^{j}{}_{j} (\omega', \omega', \vec k) 
\, 
\wt\Sigma(0, \vec k) \right]\, .
 \een
 The $-i$ and the range of the integration over $\omega'$ has 
 the following origin. While
 computing quantum corrections we shall use a Euclidean path integral with
 weight factor $e^S$. In this case the string fields will be labelled by Euclidean
 energy $\omega_E$, related to the Lorentzian energy $\omega$ via $\omega=
 i\omega_E$. The trace involved in computing the Jacobian will then involve
 $\int_{-\infty}^\infty d\omega_E/(2\pi)$, which we have expressed as
 $-i \int_{-i\infty}^{i\infty} d\omega/(2\pi)$
 The term in the exponent in \refb{e4.2xx}
 can now be interpreted as a new term in the action given by
\be\label{e5.5}
-i\, \sqrt{K\over M} \int_{-i\infty}^{i\infty} {d\omega'\over 2\pi} 
\int {d^D k\over (2\pi)^D} \, f^{j}{}_{j}(\omega', \omega', \vec k) 
\, 
\wt\Sigma(0, \vec k) \, .
\ee
Comparing this with \refb{e3.8xx} we see that this may be interpreted as 
a new contribution
to $\FF(i\vec k)$, given by
\be\label{e5.5a}
\FF_{\rm jac}= -i\,
\sqrt{K\over M} \int_{-i\infty}^{i\infty} {d\omega'\over 2\pi} \,  f^{j}{}_{j}(\omega', \omega', \vec k) 
\, ,
\ee
that needs to be added to the annulus one point function. Therefore our goal will be
to compute the functions $f^{i}{}_{j}$.

The strategy for computing $f^{i}{}_{j}$
will be as follows. We shall first write down the general effective action for $y^i$ and
$\Sigma$ consistent with the parity and time reversal symmetries 
\refb{etime}, \refb{eparity} with $\chi^i$ replaced by $y^i$ and then look for a
field redefinition \refb{e5.2gen} 
that relates the effective action of $y^i$'s to the effective action of the
$\chi^i$'s and the $\Sigma$ given in \refb{e3.8xx}.
The general 
$y^i$ and $\Sigma$ 
dependent terms in the SFT action takes the form:
\ben \label{esftexpected}
&& {K\over 2} \int {d\omega\over 2\pi} \,\omega^2 \, y^i(\omega) \, y^i(-\omega)
+ \int {d^Dk\over (2\pi)^D} \,  \,  \FF(i\vec k)  \, 
\, \wt\Sigma(0,\vec k) 
\nonumber \\ && +\ 
\int {d\omega\over 2\pi} \int {d^Dk\over (2\pi)^D} \, B^{(1)}_i(\omega,\vec k)\, 
 y^i(\omega)\, 
\wt\Sigma(-\omega, \vec k) \nonumber \\ &&
+\ {1\over 2} \int  {d\omega\over 2\pi} \,  \int {d\omega'\over 2\pi} \, 
\int {d^Dk\over (2\pi)^D} \, B^{(2)}_{ij}(\omega,\omega',\vec k)\,
y^i(\omega) \, y^j(\omega') \, \wt\Sigma(-\omega-\omega', \vec k)+\cdots\, ,
\een
where $\FF(i\vec k)$, $B^{(1)}_i(\omega,\vec k)$ and $B^{(2)}_{ij}(\omega,\omega',
\vec k)$
can be determined from the SFT action. This will be done in
section \ref{sjacob}.
Using the relation \refb{e5.2gen} between the $y^i$'s and the 
$\chi^i$'s, we can express
\refb{esftexpected} as
\ben \label{esfttrans}
&& 
{M\over 2} \int {d\omega\over 2\pi} \omega^2 \chi^i(\omega) \chi^i(-\omega)
+\int {d^Dk\over (2\pi)^D} \,  \,  \FF(i\vec k)  \, 
\, \wt\Sigma(0,\vec k) 
\nonumber \\ && +\ \sqrt{KM}\,  \int {d\omega\over 2\pi} 
\int {d^D k\over (2\pi)^D}\omega^2 \chi^i(-\omega) \, g^i(\omega,
\vec k)\, \wt\Sigma(\omega, \vec k) \nonumber \\ &&
+\  \sqrt{KM}\,  \int {d\omega\over 2\pi}
\int {d\omega'\over 2\pi} 
\int {d^D k\over (2\pi)^D}\,  \omega^2  \chi^i(-\omega) \,
f^{i}{}_{j}(\omega, \omega', \vec k)\,  \chi^j(\omega') \, 
\wt\Sigma(\omega-\omega', \vec k) 
\nonumber\\
&&
+ \sqrt{M\over K}\,
\int {d\omega\over 2\pi} \int {d^Dk\over (2\pi)^D} \, B^{(1)}_i(\omega,\vec k)
 \chi^i(\omega)
\wt\Sigma(-\omega, \vec k)
 \nonumber \\ &&
+\ {M\over 2 K} \int  {d\omega\over 2\pi} \,  \int {d\omega'\over 2\pi} \, 
\int {d^Dk\over (2\pi)^D} \, B^{(2)}_{ij}(\omega,\omega',\vec k)\,
\chi^i(\omega) \, \chi^j(\omega') \, \wt\Sigma(-\omega-\omega', \vec k)+\cdots \, .
\een
We have to compare this with \refb{e3.8xx} after expanding this in powers of
$\chi^i$:
\ben\label{e2.2cmom}
&& {M\over 2} \int {d\omega\over 2\pi} \, \omega^2\, \chi^i(\omega) \, \chi^i(-\omega) 
\nonumber \\ && +  
\int {d^Dk\over (2\pi)^D} \,  \,  \FF(i\vec k)  \, 
\bigg[ \wt\Sigma(0,\vec k) + i\, k_i \int{d\omega\over 2\pi} \chi^i(\omega) \wt\Sigma(-\omega,
\vec k) \nonumber \\ && 
-{1\over 2} k_i k_j \int{d\omega\over 2\pi} 
\int{d\omega'\over 2\pi}  \chi^i(\omega) \chi^j(\omega') \wt\Sigma(-\omega-\omega',
\vec k) 
+\cdots\bigg]\, .
\een
Comparison of \refb{esfttrans} and \refb{e2.2cmom} now gives
\ben
&& i\, k_i \, \FF(i\vec k) = \sqrt{KM} \, \omega^2\, g^i(-\omega,\vec k) 
+ \sqrt{M\over K} \, B^{(1)}_i(\omega,\vec k)  \nonumber \\
&& -{1\over 2} k_i k_j \, \FF(i\vec k)
= {\sqrt{KM}\over 2}\, \{ \omega^2 \, f^{i}{}_{j}(-\omega,\omega',\vec k)
+ (\omega')^2 \, f^{j}{}_{i}(-\omega',\omega, \vec k) \}
 + {M\over 2 K} B^{(2)}_{ij}(\omega,\omega', \vec k)\, . \nonumber \\
\een
In particular, in the second equation, after setting $\omega=-\omega'$, and 
tracing over $i,j$, we get
\be
{1\over 2} \, (\omega')^2 \left( 
f^{i}{}_{i}(\omega',\omega',\vec k) + f^{i}{}_{i}(-\omega',-\omega',\vec k)\right)
= -{1\over 2\sqrt{KM}} \left\{\vec k^2 \, \FF(i\vec k) + {M\over K} 
B^{(2)}_{ii}(-\omega',\omega',\vec k)\right\}\, ,
\ee
from which the contribution \refb{e5.5a} becomes
\be\label{e5.5abpre}
\FF_{\rm jac}={i\over 2 M}\, 
 \int_{-i\infty}^{i\infty} 
 {d\omega'\over 2\pi} \, (\omega')^{-2}\,  \left\{\vec k^2 \, \FF(i\vec k)
+ {M\over K} B^{(2)}_{ii}(-\omega',\omega',\vec k)\right\}
\, .
\ee
Due to the $1/M$ factor in the normalization,
we already have a factor of $g_s$ and so we only need to compute the tree level
contribution to $\FF(i\vec k)$ and $B^{(2)}_{ii}$ on the right hand side. 
Since we use the notation $\FF(i\vec k)$
for a general term in the effective action linear in $\Sigma$, we shall denote the
tree level contribution to $\FF$ by $\FF_0$ and rewrite the equation as
\be\label{e5.5ab}
\FF_{\rm jac}={i\over 2 M}\, 
 \int{d\omega'\over 2\pi} \, (\omega')^{-2}\,  \left\{\vec k^2 \, \FF_0(i\vec k)
+ {M\over K} B^{(2)}_{ii}(-\omega',\omega',\vec k)\right\}
\, .
\ee
We evaluate this explicitly in section \ref{sjacob}.

\sectiono{Analysis of the D0-D0-tachyon amplitude} \label{sbosonic}

In this section we shall explicitly compute the one point function of massless states
of the 26 dimensional 
bosonic string theory on an annulus with its boundary on a D0-brane
and extract a finite result for the amplitude
\refb{e1.1a} following the strategy described in the earlier sections. This will
determine the annulus contribution to $\FF(i\vec k)$, which in turn will determine
the D0-D0-tachyon amplitude via \refb{efinresult}.

\subsection{World-sheet expression for the amplitude} \label{sworld}

We take the external incoming 
closed string to carry momentum $(k^0,\vec k)$ with 
$k^2=4$. 
Therefore we can take its vertex operator to be of the form $c\bar c V$ 
with:
\be\label{evertex}
V =  e^{i k. X}=e^{-i k^0 X^0 + i\vec k.\vec X}, \qquad 
k^2=4\, .
\ee
We shall label the annulus by a complex coordinate $w$ subject to the restriction:
\be
0\le {\rm Re}(w) \le \pi, \qquad w \equiv w+2\pi i t\, .
\ee
Following the general procedure described in 
\cite{2405.19421}  and reviewed in appendix \ref{scollection}, 
we can now express the
integrand $F(x,t)$, appearing in the expression for the annulus one point
function  \refb{e1.1a} of $c\bar c V$, as,\footnote{We have not attempted
to fix the sign of this term from first principles although this could be done
following the results of \cite{2405.19421}. Instead we have chosen the sign so that
it agrees with the ones computed from Feynman diagrams. This will be seen in
section \ref{sfeyn}.}. 
\ben \label{e5.2}
F(x,t) &=& {g_s\eta_c\over 2\pi i} \times (-2\pi i) \times 2\pi 
\nonumber \\ && \left\langle \!
\left( \int_0^\pi dw b(w) + 
\int_0^\pi \bar b(\bar w) d\bar w \right)\!\! 
\left( \ointop_x dw' b(w') + \ointop_x \bar b(\bar w') d\bar w' \right) \!
c\bar c V(2\pi x)
\!\right\rangle_{\!\!A} \!,
\een
where 
\be
\eta_c\equiv {i\over 2\pi}\, .
\ee
$\ointop_x$ denotes an anti-clockwise 
contour around $x$ and $\langle\cdots\rangle_A$ denotes
unnormalized one point function on the annulus. 
The first factor of $1/2\pi i$ is the factor that accompanies the integral of $b(w)$ and
$\bar b(\bar w)$ inside the first parentheses, the second factor of $-2\pi i$ comes from the
identification $w\equiv w-2\pi i t$ so that a derivative of the transition function with 
respect to  $t$ produces a factor of $-2\pi i$ and the third factor of $2\pi$
comes from the
argument of $V$ being $2\pi x$ so that the integration measure over $x$ is $2\pi dx$. 
The $\ointop$ carry their own factors of
$\pm 1/2\pi i$.
Now the upper half plane coordinate $z$ is related to the strip coordinate $w$ via
the relation:
\be
z = e^{iw}, \qquad z\equiv e^{2\pi t} z\, .
\ee
Therefore, the expansions of $b$ and $c$ in the strip coordinate takes the form
\ben \label{eb0c0}
b(w) &=& (dz/dw)^2 b(z) = - e^{2iw}\sum_n b_n z^{-n-2} = -\sum_n b_n e^{-inw}\, ,
\nonumber \\
\bar b(w) &=& (d\bar z/d\bar w)^2 \bar b(\bar z) = - e^{-2i\bar w}\sum_n b_n \bar z^{-n-2} 
= -\sum_n b_n e^{in\bar w}\, ,
\nonumber \\
c(w) &=& (dz/dw)^{-1} c(z) = -i\, e^{-iw} \sum_n c_n z^{-n+1}
= -i \sum_n c_n e^{-inw}\, , \nonumber \\
\bar c(\bar w) &=& (d\bar z/d\bar w)^{-1} \bar c(\bar z) = i\, e^{i\bar w} \sum_n c_n \bar 
z^{-n+1}
= i \sum_n c_n e^{in\bar w}\, .
\een
After carrying out the contour
integrals we can express $F$ as:
\be \label{efxtsign}
F(x,t) = -2\pi g_s\eta_c  \left\langle \left( 2\pi \, b_0 \right)
\left(c(2\pi x) - \bar c(2\pi x) \right) V(2\pi x)
\right\rangle_A = 8\pi^2\, i\, g_s\eta_c \left\langle \left( b_0c_0\right) V(2\pi x)
\right\rangle_A,
\ee
where 
we used the fact that in order to get a non-zero result for the annulus
amplitude we must insert $b$ and $c$ zero modes.
If we denote by $\langle V(x)\rangle_N$ the normalized one point function of $V(x)$
on the annulus and by $Z(t)$ the partition function of the matter ghost CFT on the annulus
with a $b_0c_0$ insertion to soak up the ghost zero modes, then we can express
$F(x,t)$ as
\be\label{eq:f-as-one-pt-and-part}
F(x,t) =8\pi^2\, i\,  g_s\eta_c \, Z(t)\, \langle V(2\pi x)\rangle_N\, .
\ee
For $V$ of the form given in \refb{evertex}, the 
integration over the zero mode of $X^0$ produces the energy conserving delta
function $2\pi \delta(k^0)$. 
The on-shell condition now gives
\be
\vec k^2=4\, .
\ee
The one point function of $e^{i\vec k.\vec X}(2\pi x)$ can
be evaluated by using the doubling trick that 
relates it to the two point function 
\be
-\left\langle e^{ i\vec k.\vec X_R}(2\pi x) \ e^{-i\vec k.\vec X_R}(-2\pi x)\right\rangle_T
\ee 
on a torus
$T$:
\be
w\equiv w+2\pi \equiv w + 2\pi i t\, .
\ee
Here we have used the fact that the image
of $e^{i\vec k.\vec X_L}(z)$, reflected about the imaginary axis, is
$-e^{-i\vec k.\vec X_R}(-\bar z)$ for a dimension one operator.
This gives
\be\label{eq:one-pt}
\langle V(2\pi x)\rangle_N = -2\pi \delta(k^0) \left[ {2\pi \vt_1(2x|it)\over \vt_1'(0)}\right]^{-\vec k^2/2}
=-2\pi\delta(k^0){1\over 4\pi^2}  \left[ {\vt_1(2x|it)\over \vt_1'(0)}\right]^{-2}\, ,
\ee
where we used $\vec k^2=4$. 
We also have the standard expression for the D0-brane annulus partition function
\be\label{e5.14pre}
Z(t) = \eta(it)^{-24} \, \int_{-i\infty}^{i\infty} (-i) {d\omega\over 2\pi} \, e^{2\pi t \omega^2} 
={1\over 2\sqrt 2 \pi} \, t^{-1/2} \, \eta(it)^{-24} \, ,
\ee
where the $-i$ and the range of integration over $\omega$ has the same explanation
as the one given below \refb{e4.2xx}, namely that it expresses the 
integral over Euclidean energy in the Lorentzian notation.
Combining \refb{eq:one-pt} and \refb{e5.14pre},  \refb{eq:f-as-one-pt-and-part} becomes
\ben\label{e5.14}
F(x,t) &=& 2\pi \delta(k^0)\,  { g_s\eta_c'\over \sqrt 2\pi}
\,  t^{-1/2} \, \eta(it)^{-24} 
\left[ {\vt_1(2x|it)\over \vt_1'(0|it)}\right]^{-2}, \qquad \eta_c'=-i\eta_c = {1\over 2\pi}\, .
\een
Using \refb{e5.16aintro} and \refb{e5.16bintro}, we
 see that the integral over $F(x,t)$ has divergences from the 
$x\to 0$ and / or
$t\to \infty$ limit. Below we shall describe how to treat these divergences
using Feynman diagrams with open string propagators. 

\subsection{Feynman diagram analysis with tachyon propagator} \label{sfeyn}

\begin{figure}
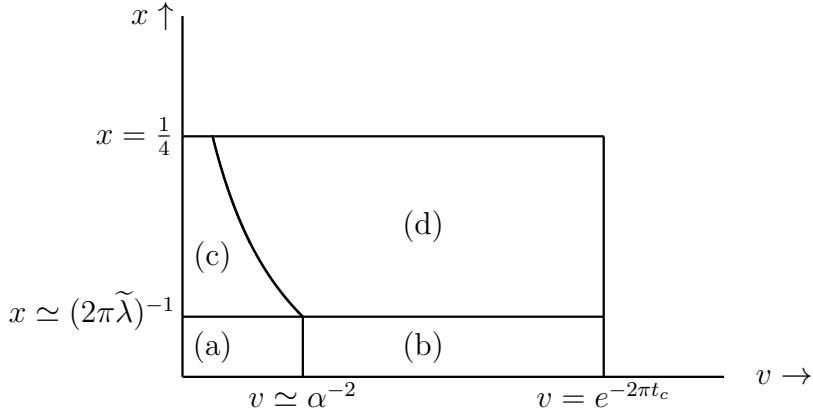

\begin{center}

\hbox{\fignine}

\vskip -.8in

\caption{The moduli space regions associated with the Feynman diagrams
shown in Fig.~\ref{figfive}(a), (b), (c) and (d), as described in 
eqs.\refb{eapp8}-\refb{eapp16}
in appendix \ref{scollection}. The boundaries between different regions 
shown here are approximate. A more precise description of these boundaries that
we use in our analysis can be found in appendix \ref{scollection}.
This figure is a reproduction of Fig.~8 of \cite{2012.11624} with minor changes.
\label{fignine}
}
\end{center}
\end{figure}

We begin by comparing the leading divergence in \refb{e5.14}  
in the $x\to 0$, $t\to\infty$ limit
with the divergences coming from Fig.~\ref{figfive}(a). For this we need to 
compute the leading divergent part of Fig.~\ref{figfive}(a) which comes from the
open string
tachyon propagating along both propagators. This has four constituents:
\begin{enumerate}
\item The open-closed interaction vertex which, in the notation of \cite{2405.19421},
is given by
\be\label{e5.18a}
\{ c\bar c  e^{ik.X}; c \}_{g=0,b=1}\, ,
\ee 
where we have set the energy carried by the open string along propagator 1 to 0 using
energy conservation.
Here $g=0$, $b=1$ implies that this involves an amplitude on a Riemann surface
of genus 0 with one boundary. Using \refb{ea4new} and taking into
account the extra minus sign mentioned below \refb{ea4new}, 
this can be expressed as an upper half plane correlation function
\be \label{e4.17}
- g_s^{1/2} 
(\eta_c)^{1/4} \, \langle c\bar c  e^{ik.X}(i)
f_0\circ c (0)\rangle_{UHP}\, ,
\ee
where $f_0$ is the conformal transformation that appears in the definition of the open-closed
interaction vertex and $f_0\circ c$ is the conformal transform of $c$ under $f_0$.
We use \refb{eapp3} to get the relation between the
upper half plane coordinate $z$ and the local coordinate $w_o$ at the open string
puncture:
\be
z = f_0(w_o), \qquad f_0(w_o)=w_o/\lambda\, .
\ee
This gives
\be
f_0\circ c(0) = \lambda \, c(0)\, .
\ee
The correlation function appearing in \refb{e4.17} has to be calculated using the
normalization given in   \refb{enormfo2}. This gives
\ben
&& \langle c(z_1) c(z_2) c(z_3) \,  e^{ik.X}(z) \rangle_{UHP} 
\nonumber \\
&=&
-K\, (z_1-z_2) (z_2-z_3) (z_1-z_3) \, 2\pi \delta(k^0) (z-\bar z)^{-\vec k^2/2}\, .
\een
Using the doubling trick we now replace $\bar c(i)$ by $c(-i)$ in \refb{e4.17}
and
use $\vec k^2=4$ to get
\be\label{e5.24}
-g_s^{1/2} 
(\eta_c)^{1/4} \, \langle c\bar c e^{ik.X}(i)
f_0\circ c(0)\rangle_{UHP} 
=-K\, 2\pi \delta(k^0) \, {i\over 2} \, \lambda \, g_s^{1/2} 
(\eta_c)^{1/4} \, .
\ee
\item The interaction vertex of three open string tachyons is given by
\ben\label{e5.24a}
\{;ce^{i\omega_1X^0},ce^{i\omega_2X^0},ce^{i\omega_3X^0}\}_{g=0,b=1} 
&=& g_s^{1/2} \eta_c^{3/4} \, \bigg[
\langle f_1\circ c e^{i\omega_1 X^0(0)} 
f_2\circ ce^{i\omega_2 X^0(0)}
f_3\circ ce^{i\omega_3 X^0(0)} \rangle \nonumber \\
&+&
\langle f_1\circ c e^{i\omega_2 X^0(0)} 
f_2\circ ce^{i\omega_1 X^0(0)}
f_3\circ ce^{i\omega_3 X^0(0)} \rangle \bigg]\, ,
\een
where $f_1,f_2,f_3$ are given in \refb{eapp2}:
\be
w_1\equiv f_1^{-1}(z) = \alpha \, {2z\over 2-z}, \qquad w_2\equiv f_2^{-1}(z) =-2 \alpha \, {1-z\over 1+z}, 
\qquad w_3\equiv f_3^{-1}(z) = \alpha \, {2\over 1-2z}\, .
\ee
This gives
\ben \label{efiexp}
&& f_1(w_1) =\alpha^{-1}w_1-{1\over 2}\alpha^{-2} w_1^2+\OO(w_1^3), \qquad
f_2(w_2) = 1 + \alpha^{-1}w_2+{1\over 2}\alpha^{-2} w_2^2+\OO(w_2^3), \nonumber \\
&&
f_3(w_3)=-\alpha \, w_3^{-1} +{1\over 2}\ ,
\een
and
\be
\{;ce^{i\omega_1X^0},ce^{i\omega_2X^0},ce^{i\omega_3X^0}\}_{g=0,b=1} 
=2\, \alpha^{3+\omega_1^2+\omega_2^2+\omega_3^2} 
\, g_s^{1/2} \, \eta_c^{3/4} \, K\, 2\pi \delta(\omega_1+\omega_2+\omega_3)\, .
\ee
\item The tachyon propagator is obtained from the kinetic term of the tachyon
\be
{K\over 2} \int {d\omega\over 2\pi} (\omega^2 +1) T(-\omega) T(\omega)\, .
\ee
This gives the propagator 
\be \label{e5.29}
- K^{-1} (\omega^2+1)^{-1} = K^{-1} \int_0^1 {dq\over q} q^{-\omega^2 -1}\, .
\ee
\end{enumerate}
Putting all these results together and taking into account a symmetry factor of $1/2$
in the loop in the Feynman diagram Fig.~\ref{figfive}(a) we get the following expression
for the contribution
\be \label{efi1}
-\int_0^1 dq_1 \int_0^1 dq_2 \int_{-i\infty}^{i\infty}
(-i)\, {d\omega'\over 2\pi} q_1^{-2} q_2^{-2-\omega^{\prime 2}}
{1\over 2} \, \times 2\, \lambda \, \alpha^{3+2\omega^{\prime 2}} \, g_s
\eta_c \, {i\over 2}\,   2\pi \delta(k^0)
\ee
We now use the leading order relation between $q_1,q_2$ and $x,t$ given in
\refb{eapp8} in the limit of large $\alpha$ and $\lambda$:
\be
e^{-2\pi t} \simeq {q_2\over \alpha^2}, \qquad 2\pi x = {q_1\over \wt\lambda},
\qquad \wt\lambda\equiv \alpha\lambda\, ,
\ee
to express \refb{efi1} as
\ben \label{e5.32}
&& -2\pi \delta(k^0) \times {i\over 2}  g_s\eta_c' \,
\int dt \int dx \int_{-i\infty}^{i\infty} 
{d\omega' \over 2\pi} \, e^{2\pi t} x^{-2} \, e^{2\pi t \omega^{\prime 2}}
\nonumber \\
&=&   {1\over 4 \sqrt 2 \pi} \,  2\pi \delta(k^0) \times g_s\eta_c' \,
\int dt \int dx \, e^{2\pi t} x^{-2} \, t^{-1/2}\, ,
\een
where we performed the gaussian integral like in \refb{e5.14pre}.
This agrees with the divergent part of \refb{e5.14} in the limit of
small $x$ and large $t$.
As shown in \cite{2012.11624}, 
the range $0\le q_1,q_2\le 1$ covers the region (a) shown in Fig.~\ref{fignine}
in the $(x,t)$ plane.

The integrals \refb{efi1} and \refb{e5.32} are of course both divergent, but the correct
prescription to deal with these divergences is to replace the integrals over $q_1$ and
$q_2$ by the left hand side of \refb{e5.29}. This gives the net contribution from the
Feynman diagrams of Fig.~\ref{figfive}(a) to be:
\be \label{eIa}
I_{(a)} = -2\pi \delta(k^0) \times {i\over 2}  g_s\eta_c \times
 \int_{-i\infty}^{i\infty} (-i)\, 
 {d\omega' \over 2\pi} \, \, \wt\lambda \, \alpha^{2+2\omega^{\prime 2}} \times 
(\omega^{\prime 2}+1+i\eps)^{-1}\, .
\ee
Note that we have included the $i\eps$ term in the open string propagator.
In the $\eps\to 0$ limit there is a pole on the imaginary $\omega'$ 
axis and we need the
$i\eps$
prescription for integrating around the singularity. As was already mentioned
earlier, we have chosen the same $i\eps$ prescription that is used for
positive mass$^2$ states. This also agrees with 
Witten's $i\eps$ prescription\cite{1307.5124}
that effectively adds $-i\eps$ to $L_0$.
This ambiguity will be absent in a tachyon free theory like type IIA or IIB 
string theory. 
Note also that under Wick rotation $\omega'\to i\omega_E'$,
$\alpha^{3+2\omega^{\prime 2}}$
becomes $\alpha^{3-2\omega_E^{\prime 2}}$ and produces exponential
suppression in the integral for large $\omega_E'$.

Regarding the one point function 
\refb{eIa} as coming from the third term in \refb{e3.8xx}, we
see that this corresponds to a contribution to $\FF(i\vec k)$ of the form
\be
\FF_{(a)}=-{i\over 2}  g_s\eta_c' \times
 \int_{-i\infty}^{i\infty} 
 {d\omega' \over 2\pi} \, \, \wt\lambda \, \alpha^{2+2\omega^{\prime 2}} \times 
(\omega^{\prime 2}+1+i\eps)^{-1}\, .
\ee
After performing the integration over $\omega'$, $\FF_{(a)}$ can be expressed as
\be
\FF_{(a)} =  {1\over 4} \,  g_s\,  \eta_c'\,  \wt\lambda\,  \left\{-i + \erfi\left(\sqrt{2\, \ln\alpha}
\right)\right\}\, ,
\ee
where $\erfi(z)$ is the
imaginary error function, defined as
\be
\erfi(z) = - {2\over \sqrt{\pi}} \, i\, \int_0^{iz} e^{-u^2} du\, .
\ee

In the analysis described above, we have only included the tachyonic
contribution but have not considered possible contribution from massless and massive
states. As already discussed in section \ref{sproblem}, the contribution from the massless
states need to be treated separately and they will not be treated using Feynman
diagrams.  However, there are also contribution from massive states.
A massive state propagating along propagator 1 will 
be accompanied by a factor of $\wt\lambda^{-1}$ and
a massive state propagating along propagator 2 will 
be accompanied by a factor of $\alpha^{-2}$
coming from the interaction vertices. For this
reason, if we drop terms carrying inverse powers of $\wt\lambda$ or $\alpha$, then the
massive state contribution can be ignored. In \cite{2012.11624}, 
where analytic expressions for the
corresponding expressions were known, this was used to avoid computing the contribution
from the massive modes, since one could systematically drop all terms containing
inverse powers of $\alpha$ or $\wt\lambda$.
But when analytic expressions are not known and the result is computed 
numerically, this procedure can miss contributions with positive power of $\alpha$
and negative power of $\wt\lambda$ or vice versa, which could give significant
contribution even in the limit of large $\alpha$ and $\wt\lambda$.
This suggests that we must also include the massive state contribution to the various
Feynman diagrams. This however will not solve the problem since 
even for determining the regions of the moduli
space covered by various Feynman diagrams,   \cite{2012.11624} used 
an approximation in which terms containing inverse powers of
$\alpha$ and / or $\wt\lambda$ were dropped. We use the same approximation in our analysis.
To compensate for this, in  section \ref{scomplete}, 
we shall use an indirect method for determining all such missing terms
at one go.
For this reason we shall not separately
discuss the computation of the contribution due to
massive open string states here.

Next we shall consider the contribution from the Feynman diagram shown in
Fig.~\ref{figfive}(c). This has two components. 
\begin{enumerate}
\item The closed-open-open three
point vertex, with the external open strings both tachyonic, will be given in the
notation of \cite{2405.19421} by, 
\be \label{e5.34a}
\{c\bar c  e^{ik.X}; c e^{i\omega' X^0}, c
e^{-i\omega' X^0}\}_{g=0,b=1}\, .
\ee
Following \refb{eapp1}, \refb{eapp15} and the sign conventions mentioned
below \refb{ea4new},  this is given
by:
\be \label{e5.34}
-2\, g_s\, \eta_c \, 
\int_{1/(2\wt\lambda)}^1 d\beta \left
\langle c\bar c  e^{ik.X}(i)\, 
\left(\,\,\ointop_{-\beta} -\ointop_\beta\right)
dz b(z) \, F_1\circ c \, e^{i\omega' X^0}(0) \, F_2\circ c \, 
e^{-i\omega' X^0}(0) \right\rangle\, ,
\ee                                                                                                                                                                                                
where  the factor of 2
accounts for an equal contribution coming from the range
$-1\le\beta\le -1/2\wt\lambda$
and, from \refb{eapp10},
\be \label{eapp10rep}
F_1(w_1) = -\beta + {4\wt\lambda^2\over 4\wt\lambda^2+1} {1+\beta^2\over
\alpha\wt\lambda} \, w_1+\OO(w_1^2),
\qquad
F_2(w_2) = \beta + {4\wt\lambda^2\over 4\wt\lambda^2+1} {1+\beta^2\over
\alpha\wt\lambda} \, w_2+\OO(w_2^2)\, .
\ee
An explanation of the overall sign in \refb{e5.34} can be given as follows. 
The $\left(\ointop_{-\beta} -\ointop_\beta\right)$ factor, with the contour integrals
in the anti-clockwise direction and $F_a$'s given as in \refb{eapp10rep}, can be
identified as $\BB_\beta$ defined in \refb{eapp0}. This is inserted to the left of
the open string vertex operator at $F_1(0)$. According to the rules described below
\refb{ea4new}, we are supposed to insert $-\BB_\beta$ and integrate $\beta$ along
the direction such that the vertex operator moves along the boundary with the
world-sheet kept to the {\em left}. In this case, as we increase $\beta$, the vertex operator
at $F_1(0)=-\beta$ moves along the real axis to the left, 
keeping the world-sheet to the {\em right}. This gives
an extra minus sign that converts $-\BB_\beta$ to $\BB_\beta$. Once we strip off
the factor of $\BB_\beta$ and the open string vertex operator at $-\beta$, we are
left with an open-closed two point function on the disk, which has an extra minus
sign according to the rules given below \refb{ea4new}. 
This explains the overall
minus sign in \refb{e5.34}.

After carrying out the contour integration  picking up residues at $\pm\beta$,
\refb{e5.34} can be expressed as
\ben
&& -2 g_s \eta_c \, \int_{1/(2\wt\lambda)}^1 d\beta  \left\{{\alpha^2\wt\lambda^2 
\over (1+\beta^2)^2} \left( 1 + {1 \over 4\wt\lambda^2}\right)^2
\right\}^{1+\omega^{\prime 2}}
\nonumber \\ &&
\left\langle c\bar c e^{ik.X}(i)
 \left(c \, e^{i\omega' X^0}(-\beta) e^{-i\omega' X^0}(\beta)
+ e^{i\omega' X^0}(-\beta) \, c\, e^{-i\omega' X^0}(\beta)\right)
\right\rangle_{UHP}\, .
\een
After evaluating the correlation function and using $\vec k^2=4$, we get
\be
-2\pi \delta(k^0) \, 2\, i\, g_s\, \eta_c \, K\,  \left\{\alpha^2\wt\lambda^2 
\left(1+{1\over 4\wt\lambda^2}\right)^2\right\}^{1+\omega^{\prime 2}}  \, 
\int_{1/(2\wt\lambda)}^1 {d\beta\over (1+\beta^2)^{1+2\omega^{\prime 2}}} (2\beta)^{2\omega^{\prime 2}}\, 
\, .
\ee
\item The contribution from the tachyon propagator is given by  \refb{e5.29}.
\end{enumerate}
Combining these results and taking into account an extra factor of 1/2 that arises
from the symmetry factor in the
open string loop and the $-i$ that accompanies integration measure over $\omega'$, 
we  get the net contribution from the Feynman diagram of
Fig.\ref{figfive}(c):
\be \label{e5.41}
-2\pi \delta(k^0) \, g_s\, \eta_c' \, i \, \int \, {d\omega'\over 2\pi} \,
\left\{\alpha^2\wt\lambda^2\left(1+{1\over 4\wt\lambda^2}\right)^2\right\}^{1+\omega^{\prime 2}} 
\int_{1/(2\wt\lambda)}^1 {d\beta\over (1+\beta^2)^{1+2\omega^{\prime 2}}} (2\beta)^{2\omega^{\prime 2}}\, 
  \int_0^1 {dq_2\over q_2} q_2^{-\omega^{\prime 2} -1}\, .
\ee
Using the result in \refb{eapp9} in the limit of small $q_2$,
\be \label{echangefigc}
\beta=\tan(\pi x), \qquad e^{-2\pi t} = q_2 \, {(1+\beta^2)^2\over 4\beta^2 \alpha^2\wt\lambda^2} \left(1+{1\over 4\wt\lambda^2}\right)^{-2}\, ,
\ee
we can express the leading term in \refb{e5.41} in the small $q_2$ limit as,
\ben
&&\hskip -.3in -2\pi \delta(k^0) \, g_s\, \eta_c' \,  i \, \pi\, \times 2\pi
\int dx \int dt \int {d\omega'\over 2\pi}
e^{2\pi t(1+\omega^{\prime 2})} {1\over \sin^2(2\pi x)}\nonumber \\
&=&  2\pi \delta(k^0) \, g_s\, \eta_c' \, {\pi \over \sqrt 2}\, \int dx \int dt 
{1\over \sin^2(2\pi x)}\,  t^{-1/2}  \, 
e^{2\pi t} \, .
\een
The integrand agrees with the leading term in \refb{e5.14} in the large $t$
limit. The integration range over $x$ and $t$ covers the region (c) in 
Fig.~\ref{fignine}\cite{2012.11624}. 

As before, the correct procedure to deal with the divergence in the large $t$ limit 
is to replace the divergent integral over $q_2$ in \refb{e5.41} by the left hand side
of \refb{e5.29}. This gives:
\be \label{e5.41new}
I_{(c)} =2\pi \delta(k^0) \, g_s\, \eta_c' \,  i \, \int{d\omega'\over 2\pi} \,
\left\{\alpha^2\wt\lambda^2\left(1+{1\over 4\wt\lambda^2}\right)^2\right\}^{1+\omega^{\prime 2}}
\int_{1/(2\wt\lambda)}^1 {d\beta\over (1+\beta^2)^{1+2\omega^{\prime 2}}} (2\beta)^{2\omega^{\prime 2}}\, 
{1\over \omega^{\prime2}+1+i\eps}\, .
\ee
This corresponds to a contribution to $\FF(i\vec k)$ of the form:
\be
\FF_{(c)}= i\, g_s\, \eta_c' \, \int{d\omega'\over 2\pi} \,
\left\{\alpha^2\wt\lambda^2\left(1+{1\over 4\wt\lambda^2}\right)^2\right\}^{1+\omega^{\prime 2}}
\int_{1/(2\wt\lambda)}^1 {d\beta\over (1+\beta^2)^{1+2\omega^{\prime 2}}} (2\beta)^{2\omega^{\prime 2}}\, 
{1\over \omega^{\prime2}+1+i\eps}\, .
\ee
After carrying out the $\omega'$ integration, we get
\be
\FF_{(c)}= -{1\over 2}\, g_s\, \eta_c' \,
\int_{1/(2\wt\lambda)}^1 {d\beta\over 4\, \beta^2} \, (1+\beta^2)\, 
\, \left\{-i+ \erfi\left(\sqrt{2 \ln\alpha + 2\ln{4\wt\lambda^2+1\over
4\wt\lambda} + 2 \ln {2\beta\over 1+\beta^2}}\right)\right\}
\, .
\ee

Next, we shall consider the contribution from the Feynman diagram of 
Fig.\ref{figfive}(b). It has three components.
\begin{enumerate}
\item The open-closed vertex is given by the same expressions as \refb{e5.24}:
\be\label{e5.24rep}
-\lambda \, g_s^{1/2} 
(\eta_c)^{1/4} \, {i\over 2}\, K\,  2\pi \delta(k^0)\, .
\ee
\item The open string propagator is given by the same expression as
\refb{e5.29} with $\omega=0$:
\be \label{e5.29reper}
- K^{-1} = K^{-1}\, \int_0^1 {dq\over q} q^{-1}\, .
\ee
\item The open string one point vertex on the annulus is given by an expression 
similar to \refb{e5.2} with the closed string vertex operator replaced by the open
string tachyon vertex operator $c$ inserted at $x=0$, the $b$ and $\bar b$ integrals
around $x$ removed and a different normalization constant that can be
read out from
\refb{eapp1}:\footnote{As in the case of \refb{e5.2}, 
the sign of \refb{e5.2open}
has been fixed
by requiring that it matches the Feynman diagram contribution of Fig.~\ref{figfive}(a)
in the large $t$ limit. In principle this could be fixed from first principles using the
result of \cite{2405.19421}.
}
\be \label{e5.2open}
 {g_s^{1/2}\eta_c^{3/4}\over 2\pi i} 
\times (-2\pi i) \times  \left\langle \left( \int_0^\pi dw b(w) + 
\int_0^\pi \bar b(\bar w) d\bar w \right)
 F_0\circ c(0)
\right\rangle_A\, .
\ee
Note that the last factor of $2\pi$ in \refb{e5.2} is absent since we do not integrate
over $x$.
$F_0$ can be read from \refb{eapp7}:
\be
w =F_0(w_o) = {3\over 2}\, i\, \alpha^{-2} - i\, \alpha^{-1} (1-\alpha^{-2})\, w_o + \cdots \, ,
\ee
where we have ignored terms of order $\alpha^{-4}$.
This gives
\be
F_0\circ c(0) = (F_0'(0))^{-1} c(w=3i\alpha^{-2}/2) = {i\alpha} \, 
\left(1 + \alpha^{-2}\right)  c(w=3i\alpha^{-2}/2)\, .
\ee
Using \refb{eb0c0} we can reduce \refb{e5.2open} to
\be
g_s^{1/2}
\eta_c^{3/4}\, 2\pi \, {\alpha}\, 
\left(1 + \alpha^{-2}\right)   \, 
\langle b_0 c_0\rangle_A = g_s^{1/2}\eta_c^{3/4}\, 2\pi
\, {\alpha}  \, \left(1 + \alpha^{-2}\right)  \, Z(t)\, .
\ee
\end{enumerate}
Multiplying all the factors we get:
\be\label{e5.49}
-2\pi \delta(k^0)\,  i\, g_s\, \eta_c\, \pi \, \left(1 + \alpha^{-2}\right) 
\, \int_{t_c}^{{1\over 2\pi}
\ln(\alpha^2-1/2)} dt\, Z(t) \, {\wt\lambda} \, 
\int dq_1\, q_1^{-2} \, ,
\ee
where the upper limit on $t$ follows from \refb{eapp14} and the lower limit $t_c$,
designed to separate out the closed string tachyon contribution, will be discussed
in section \ref{sclosed}. 
We now use \refb{eapp11},
\be
q_1 =2\pi x\wt\lambda \, (1+\alpha^{-2})\, ,
\ee
to express \refb{e5.49} as
\ben
&& -2\pi \delta(k^0)\, i\, g_s\, \eta_c\, {1\over 2} 
\, \int_{t_c}^{{1\over 2\pi}
\ln(\alpha^2-1/2)}  dt\, Z(t)  \, 
\int dx\, x^{-2} \nonumber \\
&=& 2\pi \delta(k^0)\, {1\over 4\sqrt 2 \pi} \, g_s\, \eta_c'\,  
\int_{t_c}^{{1\over 2\pi}
\ln(\alpha^2-1/2)}  dt\, \int dx\, x^{-2}
t^{-1/2} \, \eta(it)^{-24} 
 \, ,
\een
where we used \refb{e5.14pre}. This agrees with the small $x$ behaviour of
\refb{e5.14}. The integration region over $x$ and $t$ covers the region (b)
in Fig.~\ref{fignine}(b)\cite{2012.11624}

We shall remove the apparent divergence in this integral as $x\to 0$ by going back to
eq.\refb{e5.49} and replacing $\int_0^1 dq_1/q_1^2$ by $-1$ according to 
\refb{e5.29}. This gives the contribution from Fig.~\ref{figfive}(b) to be
\be \label{edefib}
I_{(b)} = 2\pi \delta(k^0)\, i\, g_s\, \eta_c\, \pi \, \left(1 + \alpha^{-2}\right) 
\, {\wt\lambda} \int_{t_c}^{{1\over 2\pi}
\ln(\alpha^2-1/2)} 
dt\, Z(t) \, .
\ee
This corresponds to a contribution of $\FF(i\vec k)$ of the form:
\be
\FF_{(b)} =  -g_s\, \eta_c'\, {1\over 2\sqrt 2 } \left(1 + \alpha^{-2}\right) 
\, {\wt\lambda} \int_{t_c}^{{1\over 2\pi}
\ln(\alpha^2-1/2)} 
dt\,  t^{-1/2} \, \eta(it)^{-24}  \, ,
\ee
where we used \refb{e5.14pre}.

Finally, 
the contribution from Fig.~\ref{figfive}(d) can now be expressed as:
\be \label{e6.2}
I_{(d)}=\int_{R'_{(d)}} F(x,t) dx dt\, ,
\ee
where $R'_{(d)}$ is the region described in \refb{eapp16} together with the
replacement of the lower limit on $t$ by $t_c$.
\ben  \label{eapp16er}
R'_{(d)} &: & {\pi\over 2}\ge 2\pi x\ge  \wt\lambda^{-1} (1-\alpha^{-2}), 
\nonumber \\
&& \hskip -.6in 
{1\over \alpha^2 \wt\lambda^2\sin^2(2\pi x)}
\left(1 + {1\over 4\wt\lambda^2}\right)^{-2} \Bigg[1+ 2 \left\{ \cot^2(2\pi x)- \wt\lambda^2 f^2\right\} \alpha^{-2}\wt\lambda^{-2}
\left(1 + {1\over 4\wt\lambda^2}\right)^{-2}\Bigg]^{-1}\le v \le e^{-2\pi t_c}\, , \nonumber \\
&& \hskip 1in f\equiv f(\tan(\pi x)), \qquad v\equiv e^{-2\pi t}\, .
\een
We have slightly changed the form of the lower bound on $v$ by dropping some
terms containing inverse powers of $\alpha$ in the expression for
$v^{-1}$, since we have been dropping these
terms anyway.
Since in $R_{(d)}$, $x$ has a lower cut-off and $t$ has an upper cut-off and a
lower cut-off, there
are no divergences from the small $x$ and /  or large $t$ region.
Using \refb{e5.14} we see that the corresponding contribution to $\FF(i\vec k)$ is:
\be\label{e5.57a}
\FF_{(d)} =   { g_s\eta_c'\over \sqrt 2\pi} \int_{R'_{(d)}} dx\,  dt \, 
\,  t^{-1/2} \, \eta(it)^{-24} 
\left[ {\vt_1(2x|it)\over \vt_1'(0|it)}\right]^{-2}\, .
\ee

\subsection{Divergences from the closed string channel} \label{sclosed}

\begin{figure}
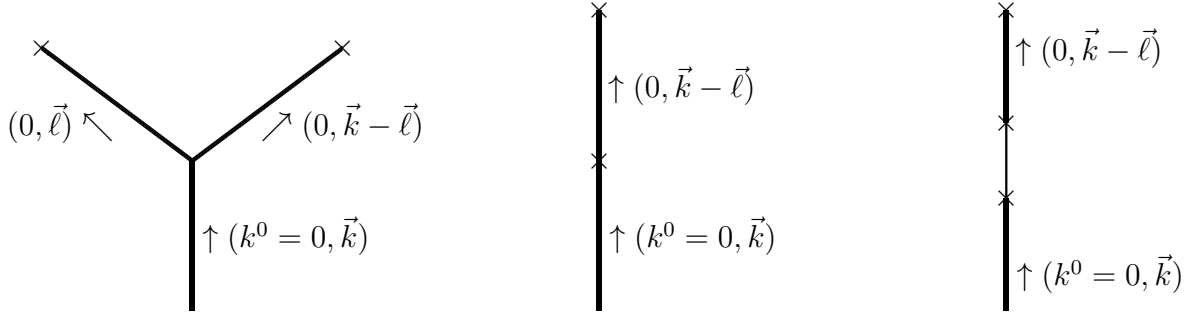

\begin{center}

\hbox{ \figtwo  \figtwob   \figtwoc 
}


\caption{This figure shows the Feynman diagram involving internal closed
string propagators for the contribution to the annulus one point function from
the small $t$ region. The $\times$'s denote
interaction vertices associated with disk amplitudes, the thick lines represent
closed strings and thin lines represent open strings. The three point interaction vertex in
the leftmost diagram is the sphere three point function of three closed strings.
\label{figtwo}
}
\end{center}
\end{figure}

We now turn to the divergences arising from the $t=0$ end of the integral.
Using the modular
transformation property
\be \label{eetamod}
\eta(it) = t^{-1/2} \eta(i/t) \, ,
\ee
\be \label{ethetaexpand}
\vt_1(2x|it)= i\, t^{-1/2} \, \exp[-4\pi x^2/t] \, \vt_1(2x / (it)|i/t)
\, ,
\ee
we can see that the integral indeed diverges in the $t\to 0$ limit. This has been
shown explicitly in appendix \ref{sb}.
These divergences arise from Feynman diagrams in open-closed
SFT containing closed string tachyons. These have been
shown in Fig.~\ref{figtwo}. 
One could carefully evaluate the contribution from the Feynman diagrams in
Fig.~\ref{figtwo} after defining the interaction vertices for off-shell closed and
open strings as in the case of Feynman diagrams of Fig.~\ref{figfive}. However,
since the closed string tachyons carry momenta $\vec\ell$ that need
to be integrated over,
we can use a shortcut based on Witten's $i\eps$ prescription\cite{1307.5124}
that is known to be equivalent to SFT evaluation of the Feynman
diagrams\cite{1610.00443}. 
For the current problem this amounts to changing 
variables from $t$ to $s=1/t$ so that the divergences arise from the $s\to\infty$
limit and then changing the upper limit of integration on $s$ to $\Lambda+i\infty$
for some large real number $\Lambda$ instead
of $\infty$. This translates to taking $t_c=(\Lambda + i\Lambda')^{-1}$ and then
taking $\Lambda'\to\infty$ limit.
We shall verify in appendix \ref{sb} that this renders the integrals finite.

\subsection{Contribution from $\tilde p, \tilde a_0, q$} \label{spqa}

\def\figsixnew{

\ifx\JPicScale\undefined\def\JPicScale{1}\fi
\unitlength \JPicScale mm
\begin{picture}(140,55.59)(0,0)
\linethickness{0.2mm}
\put(45.6,50){\circle{11.18}}

\linethickness{1mm}
\put(20,50){\line(1,0){10}}
\linethickness{0.2mm}
\put(30,50){\line(1,0){10}}

\linethickness{1mm}
\put(60,50){\line(1,0){10}}
\linethickness{0.2mm}
\put(75.5,50){\circle{11.18}}

\put(30,50){\makebox(0,0)[cc]{$\times$}}

\put(70,50){\makebox(0,0)[cc]{$\times$}}

\put(35,35){\makebox(0,0)[cc]{(e)}}

\put(75,35){\makebox(0,0)[cc]{(f)}}

\put(40,50){\makebox(0,0)[cc]{$\times$}}

\put(35,53){\makebox(0,0)[cc]{$T$}}

\put(45,58){\makebox(0,0)[cc]{$\tilde p$}}

\put(75,58){\makebox(0,0)[cc]{$\tilde p$}}

\linethickness{0.2mm}
\put(115.6,50){\circle{11.18}}

\linethickness{1mm}
\put(90,50){\line(1,0){10}}
\linethickness{0.2mm}
\put(100,50){\line(1,0){10}}

\linethickness{1mm}
\put(130,50){\line(1,0){10}}
\linethickness{0.2mm}
\put(145.5,50){\circle{11.18}}

\put(100,50){\makebox(0,0)[cc]{$\times$}}

\put(110,50){\makebox(0,0)[cc]{$\times$}}

\put(105,35){\makebox(0,0)[cc]{(g)}}

\put(145,35){\makebox(0,0)[cc]{(h)}}

\put(140,50){\makebox(0,0)[cc]{$\times$}}

\put(105,53){\makebox(0,0)[cc]{$T$}}

\put(115,58){\makebox(0,0)[cc]{$\tilde a_0$-$q$}}

\put(145,58){\makebox(0,0)[cc]{$\tilde a_0$-$q$}}

\end{picture}

}

\begin{figure}
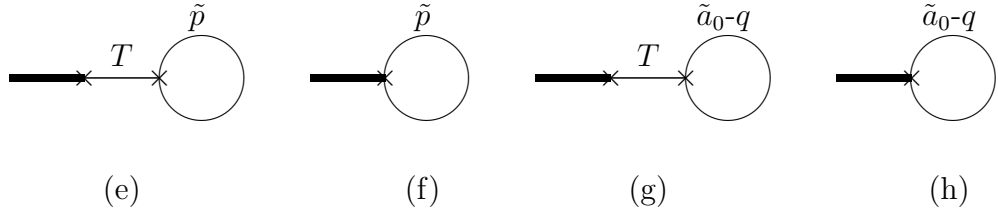

\begin{center}
\hbox{~\hskip 0in \figsixnew}
\end{center}

\vskip -1.5in

\caption{Additional Feynman diagrams contributing to the annulus one point
amplitude. Here $T$ denotes open string tachyon propagator.}  \label{figextra}
\end{figure}

In evaluating the Feynman diagrams in section \ref{sfeyn} we left out the contribution
from propagators of massless states $y^i$, $a_0$  and $p,q$ following the analysis of 
section \ref{stoy}. However, the same analysis tells us that we have to include the
contribution of the fields $\tilde p, \tilde a_0, q$ in the Feynman diagrams and treat
the integration over $y^i$ by relating it to the collective coordinates. In this section we
shall evaluate the contribution from the Feynman diagrams shown in Fig.~\ref{figfive}
when one of the internal propagators is either $\tilde p$ or the $\tilde a_0,q$ pair.

First we shall show that the propagator 1 cannot be either $\tilde p$ or the 
$\tilde a_0$-$q$ pair. For the $\tilde a_0$-$q$ pair it follows just from ghost number
conservation since only ghost number 1 states can propagate along this propagator
and $\tilde a_0$ and $q$ have ghost numbers 2 and 0 respectively.\footnote{A ghost number 1 state is needed, since the open-closed disc vertex requires a total ghost number equal to 3, the closed string vertex has ghost number 2 and there are no $B$ insertions.} For the state
$\tilde p$ the argument is a bit more subtle and was given in \cite{2012.11624}. The
essence of the argument is that for the open closed vertex the local coordinate
$w_o$ at the open string puncture is related to global coordinate $z$
in the upper half plane by the simple relation $w_o=\lambda z$ (see \refb{eapp3})
and hence the vertex operator $\p c$ for $\tilde p$ takes the same form
in $w_o$ and $z$ coordinates. Thus the evaluation of the open-closed interaction
vertex involves the ghost correlator $\langle \p c(0) c\bar c (i)\rangle$ on the upper half
plane. It is easy to see that this vanishes.\footnote{As demonstrated in \cite{2012.11624},
if we had chosen a different local coordinate at the open string puncture
of the closed-open interaction vertex then $\p c(w_o)$, expressed in the $z$ coordinate,
will have a term proportional to $c(0)$ and the contribution will not vanish.}
Therefore we only have to analyze the
Feynman diagrams shown in Fig.~\ref{figextra}.

We begin by evaluating Fig.~\ref{figextra}(e). This follows the same procedure as 
for \refb{figfive}(a) in eqs.\refb{e5.18a}-\refb{eIa} with two differences. First the 
three point vertex in \refb{e5.24a} is replaced by
\ben\label{e5.24new}
&&\{;i\p ce^{i\omega_1X^0},i\p ce^{i\omega_2X^0},ce^{i\omega_3X^0}\}_{g=0,b=1} 
\nonumber \\
&=& -g_s^{1/2} \eta_c^{3/4} \, \bigg[
\langle f_1\circ \p c e^{i\omega_1 X^0(0)} 
f_2\circ \p ce^{i\omega_2 X^0(0)}
f_3\circ ce^{i\omega_3 X^0(0)} \rangle \nonumber \\
&+&
\langle f_1\circ \p c e^{i\omega_2 X^0(0)} 
f_2\circ \p ce^{i\omega_1 X^0(0)}
f_3\circ ce^{i\omega_3 X^0(0)} \rangle \bigg]\, .
\een
Using 
\be\label{e5.59}
f\circ \p c(w) = \p c(f(w)) - {f''(w)\over (f'(w))^2} c(f(w))\, ,
\ee
and \refb{efiexp}, we get
\be
\{;i\p ce^{i\omega_1X^0},i\p ce^{i\omega_2X^0},ce^{i\omega_3X^0}\}_{g=0,b=1} 
=-2\, \alpha^{1+\omega_1^2+\omega_2^2+\omega_3^2} 
\, g_s^{1/2} \, \eta_c^{3/4} \, K\, 2\pi\, \delta(\omega_1+\omega_2+\omega_3) .
\ee
The second difference arises from the tachyon propagator $-K^{-1}(\omega^2+1)^{-1}$ 
in \refb{e5.29}
being replaced by 
$\tilde p$ propagator $1/(2K)$ following from \refb{e1.17a}. As a result, the contribution
from Fig.~\ref{figextra}(e) is obtained by multiplying the integrand in \refb{eIa} by a
factor of $\alpha^{-2} (\omega^{\prime 2}+1)/2$. This gives the contribution from this
diagram to be
\be \label{eIe}
I_{(e)} =- 2\pi \delta(k^0) \times {i\over 4}  g_s\eta_c \times
  \int_{-i\infty}^{i\infty} (-i) {d\omega' \over 2\pi} \, \, \wt\lambda \, \alpha^{2\omega^{\prime 2}} \, .
\ee
The integration over $\omega'$ is supposed to be done after the Wick rotation
$\omega'\to i\omega'_E$, and gives a finite result. The corresponding 
contribution to $\FF(i\vec k)$ is:
\be
\FF_{(e)} = -{i\over 4}  g_s\eta_c' 
  \int_{-i\infty}^{i\infty} {d\omega' \over 2\pi} \, \, \wt\lambda \, \alpha^{2\omega^{\prime 2}} 
  ={1\over 8} \, g_s\eta_c'\, \wt\lambda\, {1\over \sqrt{2\pi}} (\ln\alpha)^{-1/2}\, .
\ee

We now turn to the evaluation of Fig.~\ref{figextra}(f). This will follow the analysis
given in \refb{e5.34a}-\refb{e5.41new} for the evaluation of Fig.~\ref{figfive}(c) 
with two main
differences.
First, the closed-open-open three
point vertex, with the external open strings both tachyonic, will be replaced by, 
\be \label{e5.34aer}
\{c\bar c  e^{ik.X}; i\p c e^{i\omega' X^0}, i\p c
e^{-i\omega' X^0}\}_{g=0,b=1}\, .
\ee
Following \refb{eapp1}, \refb{eapp15} and the sign conventions below
\refb{ea4new},  this is given
by:
\be \label{e5.34ex}
-2\, \eta_c \, g_s\, 
\int_{1/(2\wt\lambda)}^1 d\beta \Bigg
\langle c\bar c  e^{ik.X}(i)\, 
\sum_{a=1}^2 \ointop_{F_a(0)} {\p F_a(w_a;\beta)\over \p\beta} 
dz b(z) 
F_1\circ \p c \, e^{i\omega' X^0}(0) \, F_2\circ \p c \, 
e^{-i\omega' X^0}(0) \Bigg\rangle\, .
\ee                                                                                                                                                                                             
Using \refb{e5.59} and \refb{eapp10}, we get
\be
F_1\circ \p c(0) = \p c(-\beta) - {h_1\over g_1^2} c(-\beta), \qquad
F_2\circ \p c(0) = \p c(\beta) - {h_2\over g_2^2} c(\beta)\, ,
\ee
and,
\be
{\p F_1\over \p\beta}=-1 + {1\over g_1} {\p g_1\over \p\beta} (z+\beta) +
\OO((z+\beta)^2), \qquad
{\p F_2\over \p\beta}=1 + {1\over g_2} {\p g_2\over \p\beta} (z-\beta)
+ \OO((z-\beta)^2)\, .
\ee
Hence
\ben
&&\sum_{a=1}^2 \ointop_{F_a(0)} {\p F_a(w_a;\beta)\over \p\beta} 
dz b(z) \, F_1\circ \p c \, e^{i\omega' X^0}(0) \, F_2\circ \p c \, 
e^{-i\omega' X^0}(0) \nonumber \\
&=& \left[ \left\{ {h_1\over g_1^2} + {1\over g_1} {\p g_1\over \p\beta}\right\}
 \left\{  \p c(\beta) - {h_2\over g_2^2} c(\beta)\right\}
- \left\{-{h_2\over g_2^2} + {1\over g_2}{\p g_2\over \p\beta}\right\}
 \left\{  \p c(-\beta) - {h_1\over g_1^2} c(-\beta)\right\}
\right] \nonumber \\ &&
\hskip 1in (g_1g_2)^{-\omega^{\prime2}}e^{i\omega' X^0}(-\beta)
e^{-i\omega' X^0}(\beta)\, . \nonumber \\
\een
Using this and the expression for the disk correlation function, \refb{e5.34ex} can be
evaluated to,
\be 
-2\pi \delta(k^0) \, 
8\,i\, K\,  \eta_c \,  g_s\, \wt\lambda^2 \, \int_{1/(2\wt\lambda)}^1 d\beta \, f(\beta)^2 \, 
{1\over 1+\beta^2} (2\beta)^{2\omega^{\prime 2}}\, 
\left\{{\alpha^2\wt\lambda^2 
\over (1+\beta^2)^2} \left(1+{1\over 4\wt\lambda^2}\right)^2 \right\}^{\omega^{\prime 2}}
\, .
\ee
The second difference arises from the tachyon propagator $-K^{-1}(\omega^2+1)^{-1}$ 
in \refb{e5.29}
being replaced by 
$\tilde p$ propagator $1/(2K)$ following from \refb{e1.17a}. 
Putting all these results together and taking into account a symmetry factor of $1/2$
in the loop in the Feynman diagram Fig.~\ref{figextra}(f) we get the following expression
for the contribution
\be
 I_{(f)} = -2\pi \delta(k^0) \, 
2\,i\,  \eta_c' \, g_s\,  \wt\lambda^2 \,  \int {d\omega'\over 2\pi} \int_{1/(2\wt\lambda)}^1 
d\beta \, f(\beta)^2 \, 
{1\over 1+\beta^2} (2\beta)^{2\omega^{\prime 2}}\, 
\left\{{\alpha^2\wt\lambda^2 
\over (1+\beta^2)^2} \left(1+{1\over 4\wt\lambda^2}\right)^2 \right\}^{\omega^{\prime 2}}
 \, .
\ee
This leads to the following contribution to $\FF(i\vec k)$:
\be
\FF_{(f)} = -2\,i\,  \eta_c' \, g_s\, \wt\lambda^2 \, \int{d\omega'\over 2\pi}
\int_{1/(2\wt\lambda)}^1 d\beta \, f(\beta)^2 \, 
{1\over 1+\beta^2} (2\beta)^{2\omega^{\prime 2}}\, 
\left\{{\alpha^2\wt\lambda^2 
\over (1+\beta^2)^2} \left(1+{1\over 4\wt\lambda^2}\right)^2 \right\}^{\omega^{\prime 2}}
 \, .
\ee
We can perform the integration over $\omega'$ after Wick rotation and arrive at the
result:
\be
\FF_{(f)} = { \eta_c' \, g_s\over \sqrt{2\pi}} \, \wt\lambda^2 \,
\int_{1/(2\wt\lambda)}^1 d\beta \, f(\beta)^2 \, 
{1\over 1+\beta^2}  {1\over  \sqrt{ \ln \alpha +
\ln{4\wt\lambda^2+1\over 4\wt\lambda}
+\ln{2\beta\over 1+\beta^2} }}
 \, .
\ee

Next we turn to Fig.~\ref{figextra}(g). For this we note that in the expansion of the
string field given in \refb{epsinew}, the vertex operators for $\tilde a_0$ and $q$ appear
in the combination:
\be
\int{d\omega\over 2\pi}\,  \left[
\tilde a_0(-\omega) {i\, \sqrt 2} \, \p c \, c \, \p X^0 + i\, q(-\omega)
\right]\, e^{i\omega X^0} \, .
\ee
The analysis  follows the same procedure as 
for Fig.~\ref{figfive}(a) in eqs.\refb{e5.18a}-\refb{eIa} with two differences. First the 
three point vertex in \refb{e5.24a} is replaced by
\ben\label{e5.24newer}
&&\{;{i \sqrt 2} \, \p c\, c\, \p X^0 e^{i\omega_1X^0}, i\, 
e^{i\omega_2X^0},ce^{i\omega_3X^0}\}_{g=0,b=1} 
\nonumber \\
&=& -\sqrt 2 g_s^{1/2} \eta_c^{3/4} \, \bigg[
\langle f_1\circ \p c c \p X^0 e^{i\omega_1 X^0}(0)
f_2\circ e^{i\omega_2 X^0}(0)
f_3\circ ce^{i\omega_3 X^0}(0) \rangle \nonumber \\
&-&
\langle f_1\circ  e^{i\omega_2 X^0} (0)
f_2\circ \p c c \p X^0 e^{i\omega_1 X^0}(0)
f_3\circ ce^{i\omega_3 X^0}(0) \rangle \bigg]\, ,
\een
where we have defined the vertex as the coefficient of the $\tilde a_0(-\omega_1)
q(-\omega_2) T(-\omega_3)$ term in the action.
The relative minus sign between the two terms reflect that $\tilde a_0$ and $q$ are
grassmann odd variables.
Using 
\ben\label{e5.59newer}
&& f\circ \p c c(0) = f'(0)^{-1}\p c c(f(0)), 
\nonumber \\ &&
 f\circ \p X^0 \, e^{i\omega X^0} (0)=
(f'(0))^{1-\omega^2} \left[
\p X^0 e^{i\omega X^0} (f(0)) + {1\over 2} \, i\, \omega {f''(0)\over (f'(0))^{2}}
e^{i\omega X^0} (f(0))\right]\, ,
\een
and \refb{efiexp}, we get
\ben
&& \{;{i \sqrt 2} \, \p c\, c\, \p X^0 e^{i\omega_1X^0},i\, 
e^{i\omega_2X^0},ce^{i\omega_3X^0}\}_{g=0,b=1} \nonumber \\
&=& \sqrt 2\, \alpha^{1+\omega_1^2+\omega_2^2+\omega_3^2} 
\, g_s^{1/2} \, \eta_c^{3/4} \, K\,  (i\omega_1)\, 2\pi\delta(\omega_1+\omega_2+\omega_3).
\een
The second difference arises from the second
tachyon propagator $-K^{-1}(\omega^{\prime 2}+1)^{-1}$ 
in \refb{e5.29}
being replaced by the
$\tilde a_0$-$q$ propagator $-i \omega_1/(\sqrt 2 K(\omega_1^2+i\eps))$ 
following from \refb{e1.17a}.  
Finally, we do not have the factor of $1/2$ from the loop since $\tilde a_0$ and $q$
are different fields.
As a result, the contribution
from Fig.~\ref{figextra}(e) is obtained by multiplying the integrand in \refb{eIa} by a
factor of $-\alpha^{-2} (\omega^{\prime 2}+1)$. This gives the contribution from this
diagram to be
\be \label{eIg}
I_{(g)} = 2\pi \delta(k^0) \times {i\over 2}  g_s\eta_c \times
 \int_{-i\infty}^{i \infty} 
 (-i)\, {d\omega' \over 2\pi} \, \, \wt\lambda \, \alpha^{2\omega^{\prime 2}} \, .
\ee
The corresponding
contribution to $\FF(i\vec k)$ is:
\be
\FF_{(g)} = {i\over 2}\, g_s\eta_c' \times
 \int {d\omega' \over 2\pi} \, \, \wt\lambda \, \alpha^{2\omega^{\prime 2}} 
 = - {1\over 4} \, g_s\eta_c'\, \wt\lambda\, {1\over \sqrt{2\pi}} (\ln\alpha)^{-1/2}\, .
\ee

We now turn to the evaluation of Fig.~\ref{figextra}(h). This will follow the analysis
given in \refb{e5.34a}-\refb{e5.41new} for the evaluation of Fig.~\ref{figfive}(c) 
with two main
differences.
First, the 
tachyon propagator $-K^{-1}(\omega^{\prime 2}+1)^{-1}$ 
is replaced by the 
$\tilde a_0(-\omega')$-$q(\omega')$ propagator 
$-i \omega'/(2\sqrt 2 K(\omega^{\prime 2}+i\eps))$ as in the
case of Fig.~\ref{figextra}(g).
Second, the
closed-open-open three
point vertex, with the external open strings both tachyonic, will be replaced by, 
\be \label{e5.34anewest}
\{c\bar c  e^{ik.X}; {i \sqrt 2} \, \p c\, c\, \p X^0 e^{i\omega' X^0}, i\, 
e^{-i\omega'X^0}\}_{g=0,b=1}\, .
\ee
Following \refb{eapp1}, \refb{eapp15} and the sign conventions below
\refb{ea4new}, this is given
by:
\ben \label{e5.34exnewest}
&& \sqrt 2 \eta_c \, 
\int_{1/(2\wt\lambda)}^1 d\beta \Bigg
\langle c\bar c  e^{ik.X}(i)\, 
\sum_{a=1}^2 \ointop_{F_a(0)} {\p F_a(w_a;\beta)\over \p\beta} 
dz b(z) \nonumber \\ &&
\left\{ F_1\circ \p c\, c\, \p X^0 e^{i\omega' X^0}(0) \, F_2\circ
e^{-i\omega' X^0}(0) 
- F_1\circ
e^{-i\omega' X^0}(0)\, F_2\circ \p c\, c\, \p X^0 e^{i\omega' X^0}(0) 
\right\}\Bigg\rangle\, . \nonumber \\
\een
We shall first evaluate the ghost part of the correlation function.                                                                                                                                                                           
Using \refb{e5.59newer} and \refb{eapp10}, we get
\ben
F_1\circ \p c\, c(0)  &=& 
\left({4\wt\lambda^2\over 4\wt\lambda^2+1} {1+\beta^2\over
\alpha\wt\lambda}\right)^{-1} 
\p c\, c(-\beta) \nonumber \\ 
F_2\circ \p c\, c(0)  &=& 
\left({4\wt\lambda^2\over 4\wt\lambda^2+1} {1+\beta^2\over
\alpha\wt\lambda}\right)^{-1} 
\p c\, c(\beta) \, .
\een
We also have
\be
{\p F_1\over \p\beta}=-1 + {2\beta\over 1+\beta^2} (z+\beta), \qquad
{\p F_2\over \p\beta}=1 + {2\beta\over 1+\beta^2} (z-\beta)\, .
\ee
Hence
\ben
&&\sum_{a=1}^2 \ointop_{F_a(0)} {\p F_a(w_a;\beta)\over \p\beta} 
dz b(z)
F_1\circ \p c\, c(0) = \left({4\wt\lambda^2\over 4\wt\lambda^2+1} {1+\beta^2\over
\alpha\wt\lambda}\right)^{-1} \left[{2\beta\over 1+\beta^2} c(-\beta) + \p c(-\beta)\right]
\nonumber \\
&&\sum_{a=1}^2 \ointop_{F_a(0)} {\p F_a(w_a;\beta)\over \p\beta} 
dz b(z)
F_2\circ \p c\, c(0) = \left({4\wt\lambda^2\over 4\wt\lambda^2+1} {1+\beta^2\over
\alpha\wt\lambda}\right)^{-1} \left[{2\beta\over 1+\beta^2} c(\beta) - \p c(\beta)\right]
\nonumber \\
\een
We now note that the relevant ghost correlators take the form:
\ben
\left\langle \left[{2\beta\over 1+\beta^2} c(-\beta) + \p c(-\beta)\right]
c(i) c(-i)\right\rangle \propto 
\left[{2\beta\over 1+\beta^2} (2i) (1+\beta^2) - 2\beta \times (2i)
\right]=0, \nonumber \\
\left\langle \left[{2\beta\over 1+\beta^2} c(\beta) - \p c(\beta)\right] c(i) c(-i)\right\rangle 
\propto \left[{2\beta\over 1+\beta^2} (2i) (1+\beta^2) - 2\beta \times (2i)
\right]=0\, .
\een
Therefore this contribution vanishes:
\be
I_{(h)}=0\, .
\ee
The corresponding contribution to $\FF(i\vec k)$ is
\be
\FF_{(h)}=0\, .
\ee

\subsection{Contribution from the Jacobian} \label{sjacob}

Finally, we shall compute the contribution to the one point function of the external
tachyon due to the Jacobian from  change of variables, as described in \refb{e5.5ab}.
\be\label{e5.5abnew}
\FF_{\rm jac}={i\over 2 M}\, 
\int_{-i\infty}^{i\infty} {d\omega'\over 2\pi} \, (\omega')^{-2}\,  \left\{\vec k^2 \, \FF_0(i\vec k)
+ {M\over K} B^{(2)}_{ii}(-\omega',\omega',\vec k)\right\}
\, .
\ee

We begin by evaluating $\FF_0(i\vec k)$. This is given by the disk one point function
of the closed string tachyon.
Using \refb{ea4new} we get
\be
2\, \pi \, \delta(k^0)\, \FF_0(i\vec k) = \eta_c^{1/2} \langle  c_0^{-} c\bar c e^{ik.X}\rangle 
= -2\, \pi \, \delta(k^0)\, K \, \eta_c^{1/2}={1\over 2} \, g_s\, M\, 2\, \pi \, \delta(k^0),
\ee
where we used $k^2=4$ for computing the matter sector correlator and used
\refb{e1.12aa} in the last step.

\begin{figure}
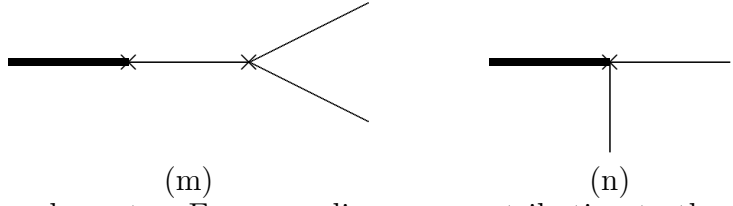

\begin{center}

\hbox{~\hskip 1in \figredef}

\vskip -1in

\caption{This figure shows two Feynman diagrams contributing to the disk amplitude with one external
closed string and two external open strings.
\label{figredef}
}
\end{center}
\end{figure}

Next, we note that $2\pi \delta(k^0)B^{(2)}_{ij}(-\omega',
\omega',\vec k) $ is the disk amplitude of the closed string
tachyon with vertex operator $c\bar c \, e^{ik.X}$ and a pair of open string states $y^i$,
$y^j$
carrying vertex operator $i\sqrt 2 \p X^i e^{ i\omega' X}$ and
$i\sqrt 2 \p X^j e^{- i\omega' X}$. This is given by the sum of the two Feynman
diagrams shown in Fig.~\ref{figredef}.

We begin with Fig.~\ref{figredef}(m). This has three components:
\begin{enumerate}
\item The open closed two point vertex with an internal tachyon is given by the same
expression as \refb{e5.24}:
\be\label{e5.24repest}
 -K\, 2\pi \delta(k^0) \, {i\over 2} \, \lambda \, g_s^{1/2} 
(\eta_c)^{1/4} \, .
\ee
\item The open string tachyon propagator is given by $-K^{-1}$.
\item The three open string interaction vertex of a tachyon, $y^i$ and $y^j$ is given by
\ben\label{e5.24aer}
&& \{;i\sqrt 2 \p X^i \, ce^{i\omega_1X^0},i\sqrt 2 \p X^j\, ce^{i\omega_2X^0},ce^{i\omega_3X^0}\}_{g=0,b=1} \nonumber \\
&=& -2 g_s^{1/2} \eta_c^{3/4} \, \bigg[
\langle f_1\circ c \p X^i e^{i\omega_1 X^0(0)} 
f_2\circ c \p X^j e^{i\omega_2 X^0(0)}
f_3\circ ce^{i\omega_3 X^0(0)} \rangle \nonumber \\
&+&
\langle f_1\circ c \p X^j \, e^{i\omega_2 X^0(0)} 
f_2\circ c \p X^i \, e^{i\omega_1 X^0(0)}
f_3\circ ce^{i\omega_3 X^0(0)} \rangle \bigg]\, .
\een
Setting $\omega_1=\omega'$, $\omega_2=-\omega'$ and $\omega_3=0$, this
evaluates to
\be
2 g_s^{1/2} \eta_c^{3/4} \, K \, \alpha^{1+2\omega^{\prime 2}} \, \delta_{ij}\, .
\ee
\end{enumerate}
Combining these contributions 
we get the net contribution to $2\pi \delta(k^0)B^{(2)}_{ij}(-\omega',
\omega',\vec k) $ from this diagram:
\be \label{e5.92}
I^{(m)}_{ij} =
i\, K\, 2\pi \delta(k^0) \,  \wt\lambda \, \alpha^{2\omega^{\prime 2}}\, g_s\, 
\eta_c \, \delta_{ij}\, .
\ee

Next we turn to Fig.~\ref{figredef}(n). This is given by an expression similar to
\refb{e5.34} with the pair of external closed string tachyons replaced by the $y^i$ and
$y^j$:
\ben \label{e5.34redef}
I_{(n)ij} &=&  -2\, g_s\, \eta_c \, 
\int_{1/(2\wt\lambda)}^1 d\beta \Bigg
\langle c\bar c  e^{ik.X}(i)\, 
\left(\ointop_{-\beta} -\ointop_\beta\right)
dz b(z) \, F_1\circ c (i\sqrt 2 \p X^i) \, e^{i\omega' X^0}(0) \nonumber \\
&&  F_2\circ c \, (i\sqrt 2
\p X^j)
e^{-i\omega' X^0}(0) \Bigg\rangle \nonumber \\
&=& 4\, g_s\, \eta_c \, \int_{1/(2\wt\lambda)}^1 d\beta 
\left[{4\wt\lambda^2\over 4\wt\lambda^2+1} {1+\beta^2\over
\alpha\wt\lambda}\right]^{-2\omega^{\prime 2}}
\nonumber \\ &&
\Bigg
\langle c\bar c (i) (c(\beta)+c(-\beta)) e^{ik.X}(i) \p X^i(-\beta) \p X^j(\beta) 
e^{i\omega' X^0}(-\beta) e^{-i\omega' X^0}(\beta) \Bigg\rangle\, .
\een
After evaluating the correlation function and dropping terms suppressed by inverse
powers of $\wt\lambda$, we get the contribution to $2\pi \delta(k^0)B^{(2)}_{ij}(-\omega',
\omega',\vec k) $
from this diagram:
\ben
I_{(n)ij} &=& 2\pi \delta(k^0) \, 
4\, i\, g_s\, \eta_c\, K \, \int_{1/(2\wt\lambda)}^1 d\beta 
\left[{4\wt\lambda^2\over 4\wt\lambda^2+1} {1+\beta^2\over
2\beta \alpha\wt\lambda}\right]^{-2\omega^{\prime 2}} 
\left[
-{1\over 8\beta^2}\, \delta_{ij} -{1\over 8}\, \delta_{ij} +{1\over 1+\beta^2}\, k_i k_j \right]
\nonumber \\
&+& 2\pi \delta(k^0) \, 
4\, i\, g_s\, \eta_c\, K \, \alpha^{2\omega^{\prime 2}}
\left[-{1\over 16\wt\lambda} \delta_{ij} + k_i k_j \, \tan^{-1}{1\over 2\wt\lambda}
\right]\, ,
\een
where the terms in the last line have been added for later convenience. We are allowed
to add these terms since they carry negative power of $\wt\lambda$ and all our
formulae\ so far allows dropping / adding such terms. For example, an exchange of 
massive states would add to \refb{e5.92} terms suppressed by powers of $\wt\lambda$
that we have not been careful to keep.

Substituting these results into \refb{e5.5abnew}, we get
\ben\label{eIjacFF}
\FF_{\rm jac} &=& {i\over 2 M} \,
\int{d\omega'\over 2\pi} \, (\omega')^{-2}\,  \Bigg[
{1\over 2} g_s M \vec k^2 
+ 25\, i\, M\,  \wt\lambda \, \alpha^{2\omega^{\prime 2}}\, g_s
\eta_c  \nonumber \\ &&
+\ 4\, i\, g_s\, \eta_c\, M \, \int_{1/(2\wt\lambda)}^1 d\beta \,
\Bigg\{{4\wt\lambda^2\over 4\wt\lambda^2+1} {1+\beta^2\over
2\beta \alpha\wt\lambda}\Bigg\}^{-2\omega^{\prime 2}} 
\Bigg\{
-{25\over 8\beta^2}-{25\over 8}+ {1\over 1+\beta^2}\, \vec k^2\Bigg\}
\nonumber \\ 
&&+ 4\, i\, g_s\, \eta_c\, M \, \alpha^{2\omega^{\prime 2}}
\left\{-{25\over 16\wt\lambda}  + \vec k^2 \, \tan^{-1}{1\over 2\wt\lambda}
\right\}\Bigg]\, .
\een
One can easily verify that the term inside the square bracket vanishes as $\omega'\to 0$
and hence the integral does not suffer from any infrared divergence. The $\wt\lambda^{-1}$
suppressed terms in the last line are important for this cancellation.

One can make the infrared finiteness of $\FF_{\rm jac}$ manifest as follows.
We first perform the $\beta$ integrals by parts to write
\ben
\FF_{\rm jac} &=& {i\over 2 M} \,
\int{d\omega'\over 2\pi} \, (\omega')^{-2}\,  \Bigg[
{1\over 2} g_s M \vec k^2 
  \nonumber \\ &&
-\ 4\, i\, g_s\, \eta_c\, M \, 2\, \omega^{\prime 2}\,
\int_{1/(2\wt\lambda)}^1 d\beta \, \Bigg\{{25\over 8\beta} 
-{25\over 8}\beta+ \vec k^2 \, \tan^{-1}\beta\Bigg\} \nonumber \\ &&
\hskip 2in \times\ \left({1\over \beta} 
- {2\beta\over 1+\beta^2}\right)
\Bigg\{{4\wt\lambda^2\over 4\wt\lambda^2+1} {1+\beta^2\over
2\beta \alpha\wt\lambda}\Bigg\}^{-2\omega^{\prime 2}}  \nonumber \\ &&
+\  i\, \pi\, \vec k^2\, 
 g_s\, \eta_c\, M \,  
\Bigg({4\wt\lambda^2\over 4\wt\lambda^2+1} {1\over
\alpha\wt\lambda}\Bigg)^{-2\omega^{\prime 2}}
\Bigg]
\, . 
\een
After using $\eta_c=i/(2\pi)$, we can write this as
\ben
\FF_{\rm jac} &=& {i\over 2 M} \,
\int{d\omega'\over 2\pi} \,  \Bigg[
-\ 8\, i\, g_s\, \eta_c\, M \, 
\int_{1/(2\wt\lambda)}^1 d\beta \, \Bigg\{{25\over 8\beta}
-{25\over 8}\beta+ \vec k^2 \, \tan^{-1}\beta\Bigg\}\nonumber \\ &&
\hskip 2in \times\
 \left({1\over \beta} 
- {2\beta\over 1+\beta^2}\right)\Bigg\{{4\wt\lambda^2\over 4\wt\lambda^2+1} {1+\beta^2\over
2\beta \alpha\wt\lambda}\Bigg\}^{-2\omega^{\prime 2}}  \nonumber \\
 &&
+i\, \pi \, \vec k^2\, 
 g_s\,  \eta_c\, M \,  {1\over \omega^{\prime 2}} \, 
 \Bigg\{\Bigg({4\wt\lambda^2\over 4\wt\lambda^2+1} {1\over
\alpha\wt\lambda}\Bigg)^{-2\omega^{\prime 2}} -1\Bigg\}
\Bigg]
\, . 
\een
The last term can be manipulated by writing $1/\omega^{\prime 2}$ as
$-d\omega^{\prime -1}/d\omega'$ and integrating by parts. This is an allowed
operation since the integral had no divergence from the $\omega'=0$ region
to start with. This gives
\ben
\FF_{\rm jac} &=& {i\over 2 M} \,
\int{d\omega'\over 2\pi} \,  \Bigg[
-\ 8\, i\, g_s\, \eta_c\, M \, 
\int_{1/(2\wt\lambda)}^1 d\beta \, \Bigg\{{25\over 8\beta} 
-{25\over 8}\beta+ \vec k^2 \, \tan^{-1}\beta\Bigg\}\nonumber \\ &&
\hskip 2in \times
\left({1\over \beta} 
- {2\beta\over 1+\beta^2}\right)
\Bigg\{{4\wt\lambda^2\over 4\wt\lambda^2+1} {1+\beta^2\over
2\beta \alpha\wt\lambda}\Bigg\}^{-2\omega^{\prime 2}}  \nonumber \\
&&
+\  4\, i\, \pi\, \vec k^2\, 
 g_s\, \eta_c\, M \, \Bigg(\ln\alpha +\ln{4\wt\lambda^2+1\over 4\wt\lambda} 
 \Bigg)
\Bigg({4\wt\lambda^2\over 4\wt\lambda^2+1} {1\over
\alpha\wt\lambda}\Bigg)^{-2\omega^{\prime 2}} 
\Bigg]
\, . 
\een
We can now carry out the $\omega'$ integration explicitly after Euclidean 
continuation, leading to,
\ben
\FF_{\rm jac} &=& -{i\over \sqrt{2\pi}} \,
 \,  \Bigg[
-\ 2\, i\, g_s\, \eta_c'\,   \int_{1/(2\wt\lambda)}^1 d\beta 
\Bigg(\ln\alpha +\ln{4\wt\lambda^2+1\over 4\wt\lambda} +\ln{2\beta\over 1+\beta^2}
 \Bigg)^{-1/2} \nonumber \\ &&
\hskip 2in \times\ \Bigg\{{25\over 8\beta} -{25\over 8}\beta+ \vec k^2 \, \tan^{-1}\beta\Bigg\}
\left({1\over \beta} 
- {2\beta\over 1+\beta^2}\right) \nonumber \\ &&
\hskip .7in +\  i\, \pi\, \vec k^2\, 
 g_s\, \eta_c'\, \Bigg(\ln\alpha +\ln{4\wt\lambda^2+1\over 4\wt\lambda} 
 \Bigg)^{1/2}
\Bigg]
\, .
\een

\sectiono{Complete annulus contribution} \label{scomplete}

Based on our results so far we conclude that the complete annulus contribution to
$\FF(i\vec k)$,  after ignoring terms with inverse
powers of $\wt\lambda$ and / or $\alpha$, is given by:
\be \label{e6.1}
\FF'\equiv \FF_{(a)} +  \FF_{(b)} +  \FF_{(c)} +  \FF_{(d)} 
+  \FF_{(e)} +  \FF_{(f)} +  \FF_{(g)} +  \FF_{(h)} + \FF_{\rm jac} \, .
\ee
However, in numerical computations below we need to work with some particular
choice of $\alpha$ and $\wt\lambda$.  Depending on this choice, either terms with a
positive power of $\alpha$ and a negative power of $\wt\lambda$ or 
vice versa that we have ignored
could become important. 
This was not a problem in the
analysis of \cite{2012.11624} since these terms are expected to cancel among 
themselves and hence in the final expression after evaluation of 
the analog of \refb{e6.1}, 
all terms containing inverse powers of $\alpha$ and / or $\wt\lambda$ were dropped.
However, unlike in the case of \cite{2012.11624}, here we shall not have
an analytic expression for the various terms from where we can explicitly drop all terms
containing inverse powers of $\alpha$ and / or $\wt\lambda$. 
In the numerical
evaluation of various terms we have to choose some large values of $\alpha$ and
$\wt\lambda$ and in that case some missing terms in the various expressions that contain
positive power of $\alpha$ and negative power of $\wt\lambda$ or vice versa can give
significant contributions. 
To overcome this difficulty, we shall choose a particular 
scaling of $\alpha$ and $\wt\lambda$, {\it e.g.} $\alpha\sim \gamma^a$, 
$\wt\lambda 
\sim \gamma^b$ for some large $\gamma$ and positive constants $a$ and $b$,
and keep all terms in  $\FF$ that scale as non-negative power of $\gamma$.
For definiteness, let us choose $a=b=1$, i.e.\ take
\be
\alpha\sim \wt\lambda \sim \gamma\, ,
\ee
with $\gamma$ being a large number. We can then evaluate $\FF$ for large
$\gamma$. By construction, \refb{e6.1} does not contain all the terms in $\FF$ that
survive in this limit, {\it e.g.} terms
proportional to $\alpha^2/\wt\lambda$ and $\wt\lambda^2/\alpha$ would be
absent in \refb{e6.1} even though they scale as $\gamma$.
We shall now describe a general procedure to determine these
missing terms.

Let us suppose that we take the expression \refb{e6.1} using the expressions for
$\FF_{(a)}$-$\FF_{\rm jac}$ as computed in the last section
and compute its variation $\delta \FF'$ 
 under arbitrary variation of $\alpha$,  $\wt\lambda$ and $f$
 without making any further approximation. If we take $\delta\alpha$ and $\delta\wt\lambda$
 to scale as $\alpha$ and $\wt\lambda$ respectively, then all terms in $\delta\FF'$
 with non-negative powers of $\alpha$ {\it and} $\wt\lambda$ must cancel since we have been
 careful to keep all such terms in our analysis. However $\delta\FF'$ may contain terms with
 negative powers of $\alpha$ {\it and / or} $\wt\lambda$ since we have dropped such terms in our
 analysis. So once we have computed $\delta\FF'$,  we  look for a term 
 $\FF_{\rm cor}$ in the form of a sum of terms, each of which carries 
 negative power of $\alpha$ {\it and /  or} $\wt\lambda$,
 whose variation explicitly cancels $\delta\FF'$. We can then add $\FF_{\rm cor}$
 to the expression for $\FF'$ given in \refb{e6.1} to
 recover $\FF$, since the latter is expected to be independent of $\alpha$, $\wt\lambda$ and
 $f(\beta)$. This procedure is completely unambiguous since the only freedom in the choice of
 $\FF_{\rm cor}$ are additive terms independent of $\alpha$, $\wt\lambda$ and $f(\beta)$, but we
 are not allowed to add such terms since they do not contain negative power of $\alpha$
 {\it or} $\wt\lambda$.

While this gives a way to recover a complete expression for $\FF$, below we shall describe the
steps to determine terms in $\FF_{\rm cor}$ that scale as non-negative power of $\gamma$,
since the other terms will be suppressed for large $\gamma$ that we shall use in our
numerical analysis.
Our first step will be to compute the change $\delta \FF'$ of
the expression for $\FF'$ under arbitrary variation of $\alpha$,  $\wt\lambda$ and $f$,
keeping all terms that scale as non-negative power of $\gamma$. 
Next we
verify that $\delta\FF'$ does not contain any term carrying non-negative power
of $\alpha$ {\it and} $\wt\lambda$. 
Therefore, the terms in $\delta\FF'$ 
will carry negative powers of $\alpha$ {\it or} $\wt\lambda$.
We shall
then explicitly add to $\FF'$ terms that will cancel this contribution. For example,
a contribution to $\delta\FF'$ of the form
\be
{\delta\wt\lambda\over \wt\lambda} {\wt\lambda^2\over \alpha^2} - 
{\delta\alpha\over \alpha} {\wt\lambda^2\over \alpha^2} 
\ee
comes from a term in $\FF'$ of the form ${1\over 2} \wt\lambda^2 \, \alpha^{-2}$.
Hence, we should add to $\FF'$ a term $-{1\over 2} \wt\lambda^2 \, \alpha^{-2}$ to
cancel this, since, if we had not made any approximation in evaluating $\FF'$, these
terms would not have been present. This procedure reduces to the procedure of
explicitly removing all terms containing inverse powers of $\alpha$ and / or $\wt\lambda$
that was used in \cite{2012.11624}.
In the end, the $\FF$ obtained this way will be the correct form of $\FF$ up to corrections
of order $\gamma^{-1}$ and we can use this for numerical computation of $\FF$ by taking
$\gamma$ to be large. This corresponds to taking $\alpha$ and $\wt\lambda$ large and
of the same order.
We could in principle find the exact expression for
$\FF$ by adding to \refb{e6.1} a term that makes $\delta \FF$ vanish exactly, but we have
not done so.

Computation of $\delta\FF'$ is tedious but straightforward after 
using the results for $\FF_{(a)}$-$\FF_{\rm jac}$ 
given in section \ref{sbosonic} and 
the result of $\omega$ integration given in
\refb{e5.14pre}. With the help
of some integration by parts and change of variable from $x$ to $\beta\equiv \tan(\pi x)$
in the expression for $\FF_{(d)}$ in \refb{e5.57a}, we get,
\ben 
\delta\FF' &=& 
-{1\over \sqrt{2\pi}} \, g_s\, \eta_c'\, {\wt\lambda^2\over \alpha^2} \, 
\int_{1/(2\wt\lambda)}^1 {d\beta\over 1+\beta^2} f(\beta)^3 \, \delta f(\beta) 
\left( \ln \alpha +\ln\wt\lambda 
+\ln{2\beta\over 1+\beta^2} \right)^{-3/2} \nonumber \\
&& - g_s\, \eta_c'\,
{1\over 4\sqrt {2\pi} }\, \wt\lambda\, (\ln\alpha)^{-3/2} \alpha^{-1}\delta\alpha 
\nonumber \\
&& +\  { g_s\, \eta_c'}\,  \sqrt{1 \over  2\pi}  \, {\delta\alpha\over \alpha} \, \int_{1/(2\wt\lambda)}^{1} {d\beta\over 1+\beta^2}\,  
\bigg[\ln\alpha +\ln{4\wt\lambda^2+1\over 4\wt\lambda} +\ln{2\beta\over 1+\beta^2}\bigg]^{-1/2} \nonumber \\ && \times \ \Bigg[{1\over 2} \, \alpha^{-2} \wt\lambda^2 \, f(\beta)^4\, \left(\ln\alpha +\ln{4\wt\lambda^2+1\over 4\wt\lambda}
+\ln{2\beta\over 1+\beta^2} \right)^{-1} \nonumber \\ &&
+{3\over 8} \, \alpha^{-2}\, \wt\lambda^2\,  f(\beta)^4 \,
\left(\ln\alpha +\ln{4\wt\lambda^2+1\over 4\wt\lambda}
+\ln{2\beta\over 1+\beta^2} \right)^{-2}
\Bigg]
\nonumber \\
&& - g_s\, \eta_c' \, {1\over 24\sqrt{2\pi}}\,  \alpha\, 
{\delta\alpha\over \wt\lambda} \,
(\ln\alpha)^{-1/2} \nonumber \\ &&
+ g_s\eta_c'\sqrt{1 \over 2\pi} {\delta\wt\lambda\over \wt\lambda}\, 
\wt\lambda^2\alpha^{-2} 
\int_{1/(2\wt\lambda)}^{1} {d\beta\over 1+\beta^2} \, \Bigg[ \ln\alpha + \ln{4\wt\lambda^2+1\over 4\wt\lambda} + 
\ln{2\beta\over 1+\beta^2}\Bigg]^{-3/2} \, f(\beta)^4 \nonumber \\ &&
\times \  \Bigg[ -  {1\over 2}
+{3\over 8} \Bigg\{ \ln\alpha + \ln{4\wt\lambda^2+1\over 4\wt\lambda} + 
\ln{2\beta\over 1+\beta^2}\Bigg\}^{-1} 
\Bigg]  \nonumber \\
&& - g_s\, \eta_c'\, {1\over 2\sqrt 2 } \,
\delta\wt\lambda\,  \left(-{1\over 3} \wt\lambda^{-2}\right)
\, \int_{t_c}^{{1\over 2\pi}
\ln(\alpha^2)} 
dt\,  t^{-1/2} \, e^{2\pi t}  
\nonumber \\ &+&
 i\, g_s\, \eta_c'\,  {\delta\wt\lambda\over 8\wt\lambda^2} \int{d\omega'\over 2\pi} \, 
\alpha^{2+2\omega^{\prime 2}} \, {1\over \omega^{\prime 2}
+1+i\eps} \, .
\een
This can be written as,
\ben
\delta\FF'
&=& g_s \eta_c' \delta \Bigg[ -{1\over 4\sqrt{2\pi}} \wt\lambda^2 \alpha^{-2} \,
\int_{1/(2\wt\lambda)}^{1} {d\beta\over 1+\beta^2}\, \left\{\ln\alpha+\ln\wt\lambda + \ln{2\beta\over 1+\beta^2}\right\}^{-3/2} \, f(\beta)^4
\nonumber \\ &&
-{i\over 8\wt\lambda} \, \int{d\omega'\over 2\pi} \alpha^{2+2\omega^{\prime 2}}
{1\over \omega^{\prime 2}+ 1+i\epsilon}
-{1\over 6\sqrt 2} \, \wt\lambda^{-1} \int_{t_c}^{{1\over 2\pi} \ln\alpha^2}
dt\, t^{-1/2} e^{2\pi t}\Bigg]\, .
\een
This suggests that we add to $\FF'$ a correction term
\ben
\FF_{\rm cor}
&=& g_s \eta_c' \Bigg[ {1\over 4\sqrt{2\pi}} \wt\lambda^2 \alpha^{-2} \,
\int_{1/(2\wt\lambda)}^{1} {d\beta\over 1+\beta^2}\, \left\{\ln\alpha+\ln\wt\lambda + \ln{2\beta\over 1+\beta^2}\right\}^{-3/2} \, f(\beta)^4
\nonumber \\ &&
+{i\over 8\wt\lambda} \, \int{d\omega'\over 2\pi} \alpha^{2+2\omega^{\prime 2}}
{1\over \omega^{\prime 2}+ 1+i\epsilon}
+{1\over 6\sqrt 2} \, \wt\lambda^{-1} \int_{t_c}^{{1\over 2\pi} \ln\alpha^2}
dt\, t^{-1/2} e^{2\pi t}\Bigg]\, .
\een
Note that in $\FF_{\rm cor}$ we are only allowed to include terms that have
either a power of $\alpha$ or a power of $\wt\lambda$ in the denominator since 
these are the types of terms that we have dropped in our analysis. In
particular, we cannot add a constant term to $\FF_{\rm cor}$.
After doing the $\omega'$ integration, we get
\ben
\FF_{\rm cor}
&=& g_s \eta_c' \Bigg[ {1\over 4\sqrt{2\pi}} \wt\lambda^2 \alpha^{-2} \,
\int_{1/(2\wt\lambda)}^{1} {d\beta\over 1+\beta^2}\, \left\{\ln\alpha+\ln\wt\lambda + \ln{2\beta\over 1+\beta^2}\right\}^{-3/2} \, f(\beta)^4
\nonumber \\ &&
- {1\over 16\wt\lambda} \, \left\{ -i + \erfi\left(\sqrt{2\ln\alpha}\right)\right\}
+{1\over 6\sqrt 2} \, \wt\lambda^{-1} \int_{t_c}^{{1\over 2\pi} \ln\alpha^2}
dt\, t^{-1/2} e^{2\pi t}\Bigg]\, .
\een
Hence, the correct form of the annulus contribution to $\FF$ is
\be \label{e6.1mod}
\FF_{\rm annulus} = \FF_{(a)} +  \FF_{(b)} +  \FF_{(c)} +  \FF_{(d)} 
+  \FF_{(e)} +  \FF_{(f)} +  \FF_{(g)} +  \FF_{(h)} + \FF_{\rm jac}+\FF_{\rm cor}
+\OO(\gamma^{-1}) \, .
\ee
We can now use this for numerical evaluation of $\FF$ by choosing some large values
of $\alpha$ and $\wt\lambda$ and some function $f(\beta)$ satisfying the boundary 
condition \refb{eflimit}.

\sectiono{Numerical evaluation of $\FF_{\rm annulus}$} \label{sec:numerics}

To obtain the numerical value of $\FF_{\rm annulus}$ summarized in~\eqref{e6.1mod}, we performed the integrations numerically using \emph{Mathematica}, and have included the corresponding code in the \texttt{arXiv} submission. We found it necessary to retain high precision (15–20 significant digits) in computing the various contributions to $\FF_{\rm annulus}$, as these often differ by several orders of magnitude. For example, for large 
$\alpha$ and $\wt\lambda$, the largest contributions, from $\FF_{(c)}$ and $\FF_{(d)}$, grow as $\alpha^2\wt\lambda^2$ and
are many orders of magnitude larger than the sum of all the terms which is independent
of $\alpha$ and $\wt\lambda$.
We note also that the imaginary part of $\FF_{\rm annulus}$ comes solely from the contributions to $\FF_{(b)}$ and $\FF_{(d)}$ from the $t$ integration contour from
$(\Lambda+i\infty)^{-1}$ to $\Lambda^{-1}$, as discussed around~\eqref{eq:iepsilon-presc}. These can be traced to internal closed string tachyon propagators.

In performing the explicit numerical calculations, we have used a family of 
functions $f_n(\beta)$ 
\be
f_n(\beta)=\frac{4 \wt\lambda ^2-3 }
{8 \wt\lambda^2 \left(1-\bigl(2 \wt\lambda\bigr)^{-n}\right)}
\left(1-\beta^n\right)
\ee
which satisfy~\eqref{eflimit}. 

\begin{figure}[t]
    \captionsetup[subfigure]{
        format=hang,            
        labelsep=space,         
        justification=raggedright,
        singlelinecheck=false   
    }

    \centering
    \begin{subfigure}{0.48\textwidth}
        \centering
        \includegraphics[width=\linewidth]{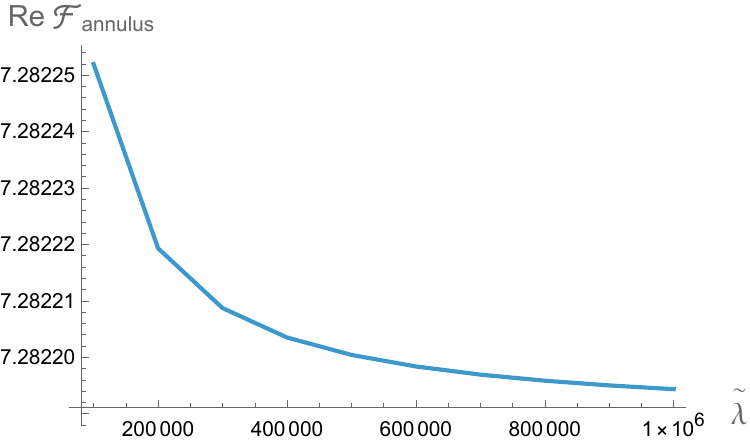}
        \caption{Real part of $\FF_{\rm annulus}$.}
        \label{fig:re-F-annulus}
    \end{subfigure}
    \hfill
    \begin{subfigure}{0.48\textwidth}
        \centering
        \includegraphics[width=\linewidth]{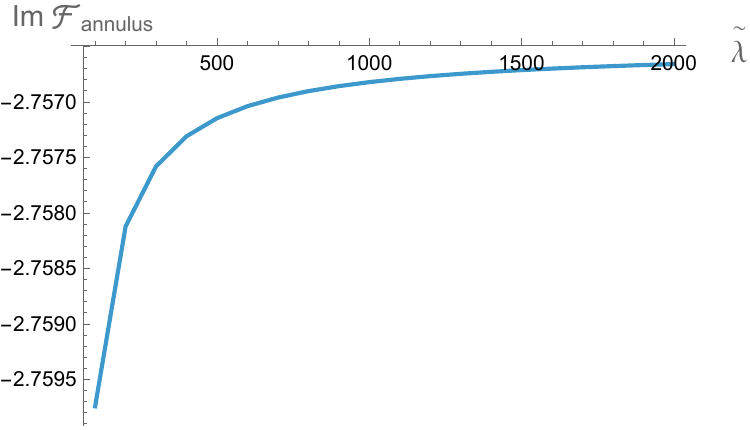}
        \caption{Imaginary part of $\FF_{\rm annulus}$.}
        \label{fig:im-F-annulus}
    \end{subfigure}

    \caption{Real and imaginary parts of $\FF_{\rm annulus}$ for $t_c=2$, $\alpha=\lambda$, and $f(x)=f_{n=1/4}(x)$.}
    \label{fig:F-annulus}
\end{figure}

In Figures~\eqref{fig:re-F-annulus} and~\eqref{fig:im-F-annulus}, we plot the real and imaginary parts of $\FF_{\rm annulus}$ for $t_c=2$, 
$\alpha=\lambda$, and $f(\beta)=f_{n=1/4}(\beta)$. We have verified numerically that $\FF_{\rm annulus}$ is independent of these choices, consistent with the general arguments presented above.
The imaginary part converges significantly more rapidly than the real part, and at $\lambda= 10^6$ we find
\be
\FF_{\rm annulus}\approx (
7.28219 - 2.75650 \, i) g_s\, \eta_c' 
\approx (1.15900-0.43871\, i) g_s\,.
\ee
Some more details of the analysis can be found in appendix \ref{app:num-res}.

\vspace{2cm}

\noindent{\bf Acknowledgement:} 
The work of A.S. was supported by the ICTS-Infosys Madhava 
Chair Professorship
and the Department of Atomic Energy, Government of India, under project no. RTI4001.
B.S. acknowledges funding support
from an STFC Consolidated Grant ‘Theoretical Particle Physics at City, University of London' ST/T000716/1 and is grateful to the CERN Theory Division for hospitality during the final stages of this project.

\appendix

\sectiono{Collection of useful results} \label{scollection}

In this appendix we shall review some of the results from SFT that we
use in our analysis.

We begin by writing down the expression for a general interaction term in open closed
SFT.  As discussed in \cite{2405.19421}, a general Riemann surface
can be described as a collection of some elementary components, {\it e.g.} a sphere
with three holes, disks around closed string punctures, semi-disks around open
string punctures etc., glued 
along closed curves $C_s$ or open curves $L_m$ beginning 
and ending on a world-sheet boundary. The information on the world-sheet moduli
$\{u^i\}$ is contained in the transition functions that relate the coordinate
$\sigma_s$ ($\sigma_m$) to the left of $C_s$ ($L_m$) and $\tau_s$ ($\tau_m$) to
the right of $C_s$ ($L_m$):
\be
\sigma_s = F_s(\tau_s, \vec u), \qquad \sigma_m = G_m(\tau_m, \vec u)\, .
\ee
We define:
\ben \label{eapp0}
\BB_i \equiv \ && 
\   \sum_s \ \biggl[ \ \ointop_{C_s} {\p F_s\over \p u^i}
d\sigma_s b(\sigma_s) + \ointop_{C_s} {\p \bar F_s\over \p u^i}
d\bar\sigma_s \bar b(\bar \sigma_s) \biggr]     \nonumber
\\
&&\hskip-10pt   +
\sum_m \ \biggl[ \ \intop_{L_m}   
{\p G_m\over \p u^i}
d\sigma_m b(\sigma_m) + \intop_{L_m}   
{\p \bar G_m\over \p u^i}
d\bar\sigma_m \bar b(\bar \sigma_m) \ 
\biggr]\, ,
\een
where $\ointop_{C_s}d\sigma_s$ and $\int_{L_m} d\sigma_m$ contain
intrinsic $1/2\pi i$ factors and
$\ointop_{C_s}d\bar\sigma_s$ and $\int_{L_m} d\bar\sigma_m$ contain
intrinsic $-1/2\pi i$ factors. For $n_c$ external closed string states
$A^c_1,\cdots,A^c_{n_c}$ and $n_o$ external open string states
$A^o_1,\cdots,A^o_{n_o}$, we now define the $p$-form on the
moduli space $M_{g,b,n_c,n_o}$ of the Riemann surface of genus $g$,
$b$ boundaries, $n_c$ closed string puncture and $n_o$ open string 
puncture\cite{2405.19421} (eqs.(3.69), (3.70)):
\ben\label{eapp1}     
&& \Omega^{(g,b,n_c,n_o)}_p (A_1^c,\cdots, A_{n_c}^c; A_1^o, 
\cdots , A_{n_o}^o ) \nonumber \\ &\equiv& 
\eta_c^{3g-3+n_c+{3\over 2} b
+{3\over 4} n_o}\, 
{1\over p!} 
du^{i_1}\wedge \cdots \wedge du^{i_p} 
\left\langle \BB_{i_1} \cdots
\BB_{i_p} \, A_1^c \cdots A_{n_c}^c; \ 
A_1^o \cdots A_{n_o}^o  
\right \rangle_{\Sigma_{g,b,n_c,n_o }}
\, ,
\een
where $\eta_c=i/(2\pi)$ and
$<\cdots>$ denotes the correlation function on the punctured Riemann surface
$\Sigma_{g,b,n_c,n_o }$. The vertex operators $A^c_i$ and $A^o_i$ are inserted
using the local coordinate system on the disks or the semi-disks on which the
corresponding puncture lies. The exception to this is the disk one point function of
closed strings for which we use
\be \label{ea4new}
\Omega^{(0,1,1,0)}_0(A^c) = \eta_c^{1/2} \langle c_0^- A^c\rangle_{0,1,1,0}\, .
\ee
Up to overall signs, the string amplitudes are obtained by integrating 
$g_s^{2g+b-2+ n_c+{1\over 2} n_o}\Omega^{(g,b,n_c,n_o)}_p$ 
over the moduli space $\MM_{g,b,n_c,n_o}$, with $p$ given by the dimension of 
the moduli space. The interaction terms of SFT are obtained by
integrating the same forms over a subspace of the moduli space $\MM_{g,b,n_c,n_o}$,
setting all the $A_i^o$'s to the open string field $\Psi_o$, all the $A_i^c$'s to the
closed string field $\Psi_c$ and dividing the result by the combinatoric factor 
$n_c!n_o!$.

For the overall signs of the amplitudes we need additional data since we have to
specify what constitutes positive integration measure in the moduli space.
For amplitudes involving purely closed strings and without boundaries, the moduli
space has a complex structure and for a complex modulus $u=u_1+i u_2$ we take
$du_1\wedge du_2$ to have positive measure and there is no additional
sign in the amplitude. In the presence of open strings or
boundaries there is no such natural choice of the sign. In \cite{2405.19421} a detailed
description of the signs of all the amplitudes were given. Here we summarize the
results that we shall need.
For the disk amplitude with one open and one
closed strings, there is an additional minus sign besides the normalization 
constants given above. Every additional open string vertex operator on the boundary,
whose location is parametrized by a modulus $u$, is accompanied by a factor of
$-\BB_u$ inserted to the immediate left of the vertex operators and the integration
over $u$ is taken to have positive measure if increasing $u$ moves the vertex operator
in a direction that keeps the world-sheet to the left\cite{2405.19421}.

Next we shall describe
the local coordinate at the open string puncture(s) in different interaction 
vertices that appear in Fig.~\ref{figfive}, following the 
conventions of \cite{2012.11624}.
We begin with the open-closed interaction vertex
that appears in Fig.\ref{figfive} (a) and (b). Representing the open-closed interaction
vertex as a correlation function on the upper half plane labelled by $z$
with the closed string inserted
at $z=i$ and the open string inserted at $z=0$, the local coordinate at the open string
puncture is taken to be\cite{2012.11624} (eq.(4.3))
\be\label{eapp3}
w_o=\lambda\, z\, ,
\ee
where $\lambda$ is an arbitrary real parameter.

For the three open string interaction vertex, described as a correlation function on the
upper half plane with the open string vertex operators inserted at $z=0$, $z=1$ and
$z=\infty$, we have the following relations between the local coordinates $w_o^{(i)}$ at
the $i$-th open string punctures and $z$\cite{2012.11624} (eq.(4.6))
\be \label{eapp2}
w_o^{(1)}= \alpha \, {2z\over 2-z}, \qquad w_o^{(2)} =-2 \alpha \, {1-z\over 1+z}, 
\qquad w_o^{(3)} = \alpha \, {2\over 1-2z}\, ,
\ee
where $\alpha$ is a real parameter that is taken to be large.

Next we consider the closed-open-open interaction vertex, represented by a 
correlation function in the upper half plane with the closed string inserted at
$i$ and the open strings inserted at $\pm\beta$ on the real line. It follows from
eq.(4.10) of \cite{2012.11624} that the local coordinates at the punctures at
$-\beta$ and $\beta$ are given respectively by $w_1=F_1^{-1}(z)$ and
$w_2=F_2^{-1}(z)$, where,
\ben
F^{-1}_1(z) &=& \alpha\wt\lambda \, {4\wt\lambda^2\over 4\wt\lambda^2+1}\, 
{z+\beta\over 1-\beta z  - \wt\lambda f(\beta)
(z+\beta)},  \nonumber \\
F^{-1}_2(z) &=& 
\alpha\wt\lambda\, {4\wt\lambda^2\over 4\wt\lambda^2+1} \, {z-\beta\over 1+\beta z + \wt\lambda f(\beta)
(z-\beta)}\, , 
\qquad \wt\lambda\equiv \alpha\, \lambda\, ,
\een
where $f(\beta)$ is an arbitrary function satisfying 
\be \label{eflimit}
f(1/2\wt\lambda)={4\wt\lambda^2-3\over 8\wt\lambda^2}, \qquad
 f(1)=0\, .
 \ee
Using
eq.(B.3), (B.10) of \cite{2012.11624} we can invert these equations as,
\ben\label{eapp10}
&& F_a(w_a,\beta)= e_a(\beta) + g_a(\beta)\, w_a + {1\over 2} h_a(\beta) \, 
w_a^2 +\OO(w_a^3), \qquad a=1,2\nonumber \\
&& e_1(\beta)=-\beta, \quad g_1(\beta)={4\wt\lambda^2\over 4\wt\lambda^2+1} {1+\beta^2\over
\alpha\wt\lambda}, \quad h_1(\beta) = -2\, \left({4\wt\lambda^2\over 4\wt\lambda^2+1}\right)^2
{\beta+\wt\lambda f\over (\alpha\wt\lambda)^2}\, (1+\beta^2)\, , \nonumber \\ 
&& e_2(\beta)=\beta, \quad g_2(\beta)={4\wt\lambda^2\over 4\wt\lambda^2+1} {1+\beta^2\over
\alpha\wt\lambda}, \quad h_2(\beta) = 2\, \left({4\wt\lambda^2\over 4\wt\lambda^2+1}\right)^2
{\beta+\wt\lambda f\over (\alpha\wt\lambda)^2}\, (1+\beta^2)\, .\nonumber \\
\een

Finally, we consider the open string one point vertex on the annulus that appears
in Fig.~\ref{figfive}(b).
We label the
points on the annulus by a complex coordinate $w$ with the restriction:
\be
0\le {\rm Re}(w) \le \pi, \qquad w \equiv w+2\pi i t\, .
\ee
Then the local coordinate $w_o$ at the open string puncture at $w=0$ is
related to $w$ via\cite{2012.11624}(eqs.(4.60), (4.68)) :
\be \label{eapp7}
w_o =2\, \alpha  {(4+3\ \alpha^{-2}) \hat z -4 + 3\, \alpha^{-2} \over
(4-\alpha^{-2}) \hat z + 4 - 7\, \alpha^{-2}}, \qquad \hat z= e^{iw}\, .
\ee
In writing this we have taken the limit of large $\alpha$ and $\wt\lambda$.

Next we shall review
the relation between the parameters $q_1$ and $q_2$ associated
with the propagators in Fig.\ref{figfive} and the parameters $x$ and $t$ labeling the
moduli space of one point function of a closed string on an annulus. 
We begin with Fig.~\ref{figfive}(a). For this diagram we have\cite{2012.11624} 
(eqs.(4.73), (4.81) together with the relation $v=e^{-2\pi t}$):
\be\label{eapp8}
e^{-2\pi t} \simeq {q_2\over \alpha^2}\left(1-{q_2\over 2 \alpha^2}\right)^{-1}, 
\qquad 2\pi x = {q_1\over \wt\lambda} \left(1-{q_2\over \alpha^2}\right)\, .
\ee
Next we consider the relation of the parameters $q_2$ and the parameter $\beta$
of the closed-open-open interaction vertex in Fig.~\ref{figfive}(c) 
with the parameters $x$ and $t$
of the annulus one point function of the closed string. We
have\cite{2012.11624} (eqs. (4.92), (4.98)):
\ben\label{eapp9}
&&\hskip -.3in  
2\pi x = 2\tan^{-1} \beta
+ {2\, u\, f} \, \wt\lambda^{-1} \,  - {1\over \beta} (1-\beta^2) \, \wt\lambda^{-2} \, u
\, ,  \qquad u \equiv q_2\, \alpha^{-2} \, \left\{ 1 + {1\over 4\wt\lambda^2}\right\}^{-2}\, ,
\nonumber \\ && 
\hskip -.3in e^{-2\pi t} = u \,  {(1+\beta^2)^2\over 4\beta^2 \wt\lambda^{2}} 
\left\{ 1 + u\, \wt\lambda^{-2}\, {1\over 2\beta^2} (1-\beta^2 - 2\beta\wt\lambda f)^2\right\}
\,.
\een
Finally, we consider the relation between the parameters $t$ of the open string one
point vertex of the annulus and the parameter $q_1$ of the open string propagator
in Fig.~\ref{figfive}(b) and the parameters  $x$ and $t$
of the annulus one point function of the closed string. The parameter $t$ is common;
so we only need to give the relation between $x$ and $q_1$. This is 
given by\cite{2012.11624} (eqs.(4.85), (4.86)):
\be\label{eapp11}
2\pi x = {q_1\over \wt\lambda}(1-\alpha^{-2})\, .
\ee

Next we shall review the range of the integration parameters associated with the
Feynman diagrams in Fig.~\ref{figfive}. First of all the $q_i$'s in each diagram are
always integrated from 0 to 1:
\be
0\le q_1\le 1, \qquad 0\le q_2\le 1\, .
\ee
In Fig.~\ref{figfive}(b), the parameter $t$ associated with the annulus one point
function of the open string is integrated over the range (eq.(4.67) of 
\cite{2012.11624}):
\be \label{eapp14}
R_{(b)} \ : \ \left(\alpha^2-{1\over 2}\right)^{-1} \le e^{-2\pi t} \le 1\, .
\ee
In Fig.~\ref{figfive}(c), the parameter $\beta$ associated with the closed-open-open
interaction vertex is integrated over the range (eq.(4.99) of \cite{2012.11624})
\be \label{eapp15}
R_{(c)} \ : \ {1\over 2\wt\lambda} \le\beta\le 1\, .
\ee
In Fig.~\ref{figfive}(d), the range of $x$ and $\beta$ are (eq.(4.102) of \cite{2012.11624}):
\ben  \label{eapp16}
R_{(d)} &: & {\pi\over 2}\ge 2\pi x\ge  \wt\lambda^{-1} (1-\alpha^{-2}), 
\nonumber \\
&& \hskip -.3in 
{1\over \alpha^2 \wt\lambda^2\sin^2(2\pi x)}
\left(1 + {1\over 4\wt\lambda^2}\right)^{-2} \Bigg[1- 2 \left\{ \cot^2(2\pi x)- \wt\lambda^2 f^2\right\} \alpha^{-2}\wt\lambda^{-2}
\left(1 + {1\over 4\wt\lambda^2}\right)^{-2}\Bigg]\le v <1\, , \nonumber \\
&& \hskip 1in f\equiv f(\tan(\pi x)), \qquad v\equiv e^{-2\pi t}\, .
\een

\sectiono{Closed string tachyons} \label{sb}

As already discussed in section \ref{sbosonic}, the integrands of $I_{(b)}$
and $I_{(d)}$ diverge as $t\to 0$, rendering the integrals divergent if we take
the lower limit of $t$ to be zero. In this appendix we shall verify that Witten's
$i\eps$ prescription, discussed at the end of section \ref{sbosonic}, makes these
integrals finite.

We begin with $I_{(d)}$ for which the integrand is
\be\label{e5.14repa}
F(x,t) = 2\pi \delta(k^0)\,  { g_s\eta_c'\over \sqrt 2\pi}
\,  t^{-1/2} \, \eta(it)^{-24} 
\left[ {\vt_1(2x|it)\over \vt_1'(0|it)}\right]^{-2}\, .
\ee
To study its behaviour at small $t$, we make a change
of variable from $t$ to $s$:
\be
s = {1\over t}\, ,
\ee
so that $t\to 0$ corresponds to $s\to\infty$. The annulus can now be regarded as
having circumference $2\pi/s$ and width $\pi$, but by scaling the coordinates
by $s$ we can also describe this as having circumference $2\pi$ and
width $\pi s$. Thus physically $\pi s$ represents the distance over which the 
closed string propagates.
We now use the modular transformation properties
\refb{eetamod} and \refb{ethetaexpand} to write
\be
\eta(it) = t^{-1/2} \eta(i/t) = s^{1/2} \eta(is)\, ,
\ee
\be
\vt_1(2x|it)= i\, t^{-1/2} \, \exp[-4\pi x^2/t] \, \vt_1(2x / (it)|i/t)
= i\, s^{1/2} \, \exp[-4\pi x^2 s] \, \vt_1(-2ixs|is)\, .
\ee
Using the product representations 
\refb{e5.16aintro}, \refb{e5.16bintro} we now get, 
\be \label{eetaexp}
(\eta(it))^{-24}  = s^{-12} \, e^{2\pi s} \prod_{n=1}^\infty (1 - e^{-2\pi n s})^{-24}\, ,
\ee
\be
\vt_1(-2ixs|is) = 2 \, i\, e^{-\pi s/4}\, \sinh(2\pi x s)\, \prod_{n=1}^\infty 
\{ (1- e^{-2\pi  n s}) (1 - 2\, e^{-2\pi n s}\,  \cosh(4\pi x s) + e^{-4\pi  n s})\}\, .
\ee
Using this in \refb{e5.14repa}, we see that $F(x,t)$ 
has divergence from the $s\to\infty$ limit due to the $e^{2\pi s}$
factor in \refb{eetaexp}. As anticipated, the origin of this can be traced to the closed string tachyon.
The rest of the factor can expanded in a power series in $e^{-2\pi s}$, $e^{-2\pi sx}$
and $e^{-2\pi s(1-x)}$
and gives a convergent expansion.

Now suppose that we express the integral over $x$ and $t$ as an integral over $x$ and
$s$ and take the $s$ integration contour to run along the positive real axis to some large
number $\Lambda$ and then turn the contour parallel to the imaginary axis towards
$\Lambda+i\infty$. Then the offensive $e^{2\pi s}$ factor becomes oscillatory, and
the integral converges due to the $s^{-12}$ factor in \refb{eetaexp}.
This is Witten's $i\eps$ 
prescription\cite{1307.5124}, but this is equivalent to treating the 
divergences using the Feynman diagrams of open closed SFT shown
in Fig.~\ref{figtwo}.

Similar analysis can be done for the integrand of $I_{(b)}$ given in \refb{edefib}. 
In this case the $t$ dependent part of the integrand is $Z(t)\propto \eta(it)^{-24}$
and hence is given by \refb{eetaexp}. Arguments identical to the ones given above
show that this integral diverges if the upper limit of $s$ integration is taken to be $\infty$,
but we can get a convergent result by taking the upper limit to be $\Lambda+
i\Lambda'$ for large positive $\Lambda,\Lambda'$ and then taking $\Lambda'$ to
$\infty$.

\sectiono{Numerical results}\label{app:num-res}

\begin{table}[htbp]
\centering
\begin{tabular}{|r|r|r|}
\hline
$\wt\lambda=5\alpha$ & $\textrm{Re}\,\FF_{\rm annulus}$ & $\textrm{Im}\,\FF_{\rm annulus}$ \\
\hline
100000  & 7.284015676539 & -2.7565011017855 \\
200000  & 7.28307221834  & -2.75649947586   \\
300000  & 7.2827668799   & -2.75649893385   \\
400000  & 7.2826168249   & -2.7564986628    \\
500000  & 7.282527888    & -2.7564985002    \\
600000  & 7.282469155    & -2.756498392     \\
700000  & 7.282427526    & -2.756498314     \\
800000  & 7.282396506    & -2.756498256     \\
900000  & 7.28237251     & -2.756498211     \\
1000000 & 7.28235341     & -2.756498175     \\
\hline
\end{tabular}
\caption{Numerical values for $\wt\lambda$, $\textrm{Re}\,\FF_{\rm annulus}$, and $\textrm{Im}\,\FF_{\rm annulus}$, with $\alpha=\wt\lambda/5$.}
\label{tab:lambda-trim-1/5}
\end{table}
\begin{table}[htbp]
\centering
\begin{tabular}{|r|r|r|}
\hline
$\wt\lambda=\alpha$ & $\textrm{Re}\,\FF_{\rm annulus}$ & $\textrm{Im}\,\FF_{\rm annulus}$ \\
\hline
100000  & 7.28225194597 & -2.75650110179 \\
200000  & 7.2822192953  & -2.7564994759  \\
300000  & 7.282208705   & -2.756498934   \\
400000  & 7.28220349    & -2.756498663   \\
500000  & 7.28220040    & -2.75649850    \\
600000  & 7.28219836    & -2.75649839    \\
700000  & 7.2821969     & -2.75649831    \\
800000  & 7.2821958     & -2.75649826    \\
900000  & 7.2821950     & -2.7564982     \\
1000000 & 7.2821943     & -2.7564982     \\
\hline
\end{tabular}
\caption{Numerical values for $\wt\lambda$, $\textrm{Re}\,\FF_{\rm annulus}$, and $\textrm{Im}\,\FF_{\rm annulus}$, with $\alpha=\wt\lambda$.}
\label{tab:lambda-trim-1}
\end{table}
\begin{table}[htbp]
\centering
\begin{tabular}{|r|r|r|}
\hline
$\wt\lambda=\alpha/5$ & $\textrm{Re}\,\FF_{\rm annulus}$ & $\textrm{Im}\,\FF_{\rm annulus}$ \\
\hline
100000  & 7.282186213 & -2.7565011018 \\
200000  & 7.28218737  & -2.756499476  \\
300000  & 7.2821878   & -2.75649893   \\
400000  & 7.2821880   & -2.7564987    \\
500000  & 7.2821881   & -2.7564985    \\
600000  & 7.282188    & -2.7564984    \\
700000  & 7.282188    & -2.756498     \\
800000  & 7.282188    & -2.756498     \\
900000  & 7.282188    & -2.756498     \\
1000000 & 7.28219     & -2.756498     \\
\hline
\end{tabular}
\caption{Numerical values for $\wt\lambda$, $\textrm{Re}\,\FF_{\rm annulus}$, and $\textrm{Im}\,\FF_{\rm annulus}$, with $\alpha=5\wt\lambda$.}
\label{tab:lambda-trim-5}
\end{table}

In this appendix we summarise the numerical results for estimating $\FF_{\rm annulus}$. An ancilliary Mathematica notebook is attached to this submission~\cite{mma-supp}. Here we present the numerical estimates for  $1\times 10^5\le\wt\lambda\le 10^6$, in the three cases of $\alpha=\wt\lambda/5$, $\wt\lambda$, $5\wt\lambda$. These are presented in Tables~\ref{tab:lambda-trim-1/5},~\ref{tab:lambda-trim-1} and~\ref{tab:lambda-trim-5}, respectively. The number of decimal places presented in the tables varies because each value is printed with the precision returned by Mathematica; we avoid zero-padding to prevent implying spurious accuracy. 

We also fitted the data given in each of these tables to a trial function
\be
c_0+ c_1 \, \wt\lambda^{-1} (\ln\wt\lambda)^{3/2} +  c_2 \, \wt\lambda^{-1}
 (\ln\wt\lambda)^{1/2} 
 + c_3\,  \wt\lambda^{-1} (\ln\wt\lambda)^{-1/2}\, ,
\ee
with the $c_i$'s allowed to be different for different tables. However we find that for
each of the three tables, we get
\be
c_0 = 7.28219 - 2.75650 \, i\, .
\ee
Hence we can take this to be the result for $\wt\lambda\to\infty$.

\end{document}